\documentclass{pasj00}
\usepackage{epsf}
\draft

\newcommand{\RG}{{\rm\scriptscriptstyle RG}}
\newcommand{\LN}{{\rm\scriptscriptstyle LN}}
\newcommand{\num}{{\nu_{\rm m}}}
\newcommand{\nuf}{{\nu_{\rm f}}}
\newcommand{\nus}{{\nu_{\sigma}}}
\newcommand{\deltam}{{\delta_{\rm mass}}}
\newcommand{\sigmamm}{{\sigma_{\rm mm}}}
\SetRunningHead{Hikage et al.}
{3D Genus of SDSS Early Data Release Galaxies}
\Received{2002 July 18}
\Accepted{2002 August 21}
\begin{document}
\title{Three--Dimensional Genus Statistics 
of Galaxies \\ in the SDSS Early Data Release}
\author{%
Chiaki \textsc{Hikage}\altaffilmark{1}, 
Yasushi \textsc{Suto},\altaffilmark{1},
Issha \textsc{Kayo}\altaffilmark{1}, 
Atsushi \textsc{Taruya}\altaffilmark{1}, \\
Takahiko \textsc{Matsubara}\altaffilmark{2},
Michael S. \textsc{Vogeley}\altaffilmark{3}, 
Fiona \textsc{Hoyle}\altaffilmark{3}, \\
J. Richard \textsc{Gott III}\altaffilmark{4},   
Jon \textsc{Brinkmann}\altaffilmark{5},  
for the SDSS collaboration}

\altaffiltext{1}{Department of Physics, School of Science, 
University of Tokyo, Tokyo 113-0033}
\altaffiltext{2}{Department of Physics and Astrophysics, 
Nagoya University, Chikusa, Nagoya 161-8602}
\altaffiltext{3}{Department of Physics, Drexel University, 3141
Chestnut Street, Philadelphia, PA 19104, USA}
\altaffiltext{4}{Princeton University Observatory, Peyton Hall,
Princeton, NJ 08544, USA}
\altaffiltext{5}{Apache Point Observatory, P.O.Box 59, Sunspot NM
88349-0059, USA}
\email{hikage@utap.phys.s.u-tokyo.ac.jp, suto@phys.s.u-tokyo.ac.jp, 
kayo@utap.phys.s.u-tokyo.ac.jp, \\ ataruya@utap.phys.s.u-tokyo.ac.jp,
taka@a.phys.nagoya-u.ac.jp, \\
hoyle@venus.physics.drexel.edu, vogeley@drexel.edu,
jrg@astro.princeton.edu, jb@apo.nmsu.edu}
\maketitle
\begin{abstract}
  We present the first analysis of three-dimensional genus statistics
  for the SDSS EDR galaxy sample. Due to the complicated survey volume
  and the selection function, analytic predictions of the genus
  statistics for this sample are not feasible, therefore we construct
  extensive mock catalogs from N-body simulations in order to compare
  the observed data with model predictions.  This comparison allows us
  to evaluate the effects of a variety of observational systematics on
  the estimated genus for the SDSS sample, including the shape of the
  survey volume, the redshift distortion effect, and the radial
  selection function due to the magnitude limit.  The observed genus for
  the SDSS EDR galaxy sample is consistent with that predicted by
  simulations of a $\Lambda$-dominated spatially-flat cold dark matter
  model. Standard ($\Omega_0=1$) cold dark matter model predictions do
  not match the observations. We discuss how future SDSS galaxy samples
  will yield improved estimates of the genus.
\end{abstract}

\section{Introduction}

Characterizing the large-scale distribution of galaxies and clusters is
critical for understanding the formation and evolution of this structure
as well as for probing the initial conditions of the universe
itself. The most widely used statistical measure for this purpose is the
two-point correlation function (2PCF).  This statistic is easily
estimated, is simply parameterized, and is well-studied.  The observed
correlation function is approximately a power law over a finite range of
scales (Totsuji \& Kihara 1969), and thus can be expressed by two
numbers, the power-law slope and the correlation length. This
correlation function statistic has been successfully applied in the
cosmological context for more than 30 years, thus its behavior is well
understood theoretically. Given a set of cosmological parameters, one
can predict the corresponding mass 2PCF $\xi(r,z)$ on a scale $r$ at a
redshift $z$ using accurate fitting formulae (e.g., Hamilton et
al. 1991; Peacock \& Dodds 1996).  In principle, the spatial biasing of
luminous objects relative to the underlying dark matter may invalidate
the straightforward comparison between observations and theoretical
predictions.  However, the SDSS galaxy sample which we analyze is
consistent with a (practically) scale-independent linear bias on scales
of $(0.2 - 4)h^{-1}$Mpc if one adopts the currently popular
$\Lambda$-dominated spatially-flat cold dark mater (LCDM) model (Kayo et
al. 2002).

While the 2PCF and its Fourier transform the power spectrum are the most
convenient and useful cosmological statistics, they completely ignore
information about the correlations of the phases of the density
fluctuations in $k$-space.  In contrast, the topology of large-scale
structure, as measured by the genus statistic (Gott, Melott, \&
Dickinson 1986) is strongly sensitive to these phase correlations.  One
of the Minkowski functionals (Mecke, Buchert \& Wagner 1994; Schmalzing
\& Buchert 1997) is closely related to the genus.  Other statistics that
quantify phase correlations in the data include higher-order n-point
correlation functions (e.g., Peebles 1980 and references therein),
percolation analysis (Shandarin 1983), minimal spanning trees (Barrow,
Bhavsar \& Sonoda 1985), and void statistics (White 1979).  We focus on
the genus mainly because its behavior is well-understood theoretically;
if the primordial fluctuations are Gaussian, then the genus in the
linear regime has an exact analytic expression (see
eq. [\ref{eq:genus_rd}] below). The theoretical prediction for a
Gaussian random field makes it clear that only the amplitude of the
genus depends on the 2PCF, while higher order correlations determine the
shape as well as affecting the amplitude.  Thus, the shape of the genus
statistic plays a complementary role to the 2PCF. In addition to the
linear theory predictions, a perturbative expression for the genus in
the weakly nonlinear regime has been obtained by Matsubara (1994) and
the log-normal model is shown to be a good empirical approximation in
the strongly nonlinear regime (e.g., Coles \& Jones 1991; Hikage, Taruya
\& Suto 2002).

Previous to the SDSS, investigations of the 3D genus statistic for
galaxy redshift surveys include Gott et al. (1989), Park, Gott \& da
Costa (1992), Moore et al. (1992) Rhoads et al. (1994), Vogeley et al.
(1994), and Canavezes et al. (1998). As suggested by Melott (1987), the
2D variant of this statistic has been estimated for a variety of data
sets by Coles \& Plionis (1991), Plionis, Valdarnini \& Coles (1992),
Park et al. (1992), Colley (2000), Park, Gott, \& Choi (2001), and
Hoyle, Vogeley \& Gott (2002a). These investigations generally indicate
consistency with the hypothesis that the initial perturbations were
Gaussian in nature. Some departures from Gaussianity have been
suggested, but the statistical significance of the results was low due
to the small size of the available data sets.  Hoyle et al. (2002b) find
weak evidence for variation in the genus with galaxy type in the SDSS
using the 2D genus statistic.

In the present paper, we describe the first analysis of the 3D genus of
the SDSS early data release galaxy sample.  We evaluate a variety of
observational effects using mock catalogs from N-body simulations such
as the shape of the survey volume, the redshift distortion effect, and
the radial selection function due to the magnitude limit. To within the
uncertainties for this preliminary sample, we find that the LCDM model
reasonably reproduces the observed shape and amplitude of the genus of
SDSS galaxies. A complementary analysis of the 2D genus for the SDSS
galaxies is presented separately in Hoyle et al. (2002b).  Other
analyses of early SDSS data include measurement of the power spectrum
(Szalay et al. 2002; Tegmark et al. 2002; Dodelson et al. 2002),
correlation function (Zehavi et al. 2002; Connolly et al. 2002; Infante
et al. 2002), and higher-order moments (Szapudi et al. 2002).

The outline of the paper is as follows. In section 2 we describe the
SDSS Early Data Release sample that we analyze. To analyze systematic
effects we use mock samples drawn from N-body simulations, which are
described in section 3. Section 4 describes the genus statistic and
the method of estimation. Observational systematics are analyzed in
section 5. Genus results for the SDSS EDR sample are presented in
section 6. Section 7 presents our conclusions.

\section{SDSS galaxy data}

The early data release sample of the SDSS includes five-color CCD
imaging and moderate-resolution spectroscopy of galaxies in Northern and
Southern equatorial stripes of width $2.5^\circ$ in declination and in
another stripe that overlaps the SIRTF First Look Survey.  York et
al. (2000) provide an overview of the SDSS. Stoughton et al.  (2002)
describe the early data release (EDR) and details about the
measurements.  Technical articles providing details of the SDSS include
description of the photometric camera (Gunn et al. 1998), photometric
analysis (Lupton et al. 2002), the photometric system (Fukugita et al.
1996; Hogg et al. 2001; Smith et al. 2002), astrometric calibration
(Pier et al. 2002), selection of the galaxy spectroscopic samples
(Strauss et al. 2002; Eisenstein et al.  2001), and spectroscopic tiling
(Blanton et al. 2002).

The present analysis is based on the Northern stripe, which covers
$145^\circ.89 < \alpha < 234^\circ.57$ (excluding three unobserved gap
regions; $152^\circ.46 <\alpha <156^\circ.02$,
$167^\circ.03<\alpha<171^\circ.88$, $201^\circ.57<\alpha<205^\circ.36$)
and the Southern stripe, which covers
$351^\circ.20<\alpha<55^\circ.49$. We form a sub-sample of galaxies with
$14.5 < r^\ast < 17.6$, after correction for Galactic reddening using
the maps of Schlegel, Finkbeiner, \& Davis (1998).  (The SDSS wavelength
band used for galaxy target selection is denoted by $r'$, but we use
$r^\ast$ to indicate apparent magnitudes using preliminary photometric
calibration.)  The net sample includes $19710$ and $15977$ galaxies in
the Northern and the Southern stripes, respectively. The resulting
apparent-magnitude limited sample has a different mix of galaxy types at
different redshifts. We do not consider the effect of this redshift
dependence in the following analysis.  Note, however, that Hoyle et
al. (2002b) find a small difference of the two-dimensional genus for the
reddest and bluest galaxies.

We estimate the genus for the three-dimensional galaxy distribution in
redshift space. For this purpose, we use the comoving distance computed
from the observed redshift $z$ (after correction for the Local Group
motion) of each galaxy:
\begin{equation}
\label{eq:rz}
r(z)=\int_0^{z}\frac{1}
{\sqrt{\Omega_0(1+z^\prime)^3+(1-\Omega_0-\lambda_0)(1+z^\prime)^2
+\lambda_0}}dz^\prime ,
\end{equation}
where $\Omega_0$ is the matter density parameter and $\lambda_0$ is the
dimensionless cosmological constant.  Because of this mapping, the
density fields, and thus the genus constructed from the observed galaxy
distribution, are dependent on the assumed set of cosmological
parameters discussed below.

\section{Mock samples from N-body simulations}

To quantitatively evaluate various observational effects on the genus
statistics such as the shape of survey volume, the redshift distortion
effect, and the radial selection function, we generate various different
mock samples which are summarized in Table \ref{tab:data}.  All the mock
catalogs are constructed from a series of the P${}^3$M N-body
simulations by Jing \& Suto (1998) which employ $256^3$ particles in a
$(300h^{-1}{\rm Mpc})^3$ periodic comoving box.  These simulations use
Gaussian initial conditions and the cold dark matter (CDM) transfer
function (Bardeen et al. 1986).  We neglect the light-cone effect
(Hamana, Colombi \& Suto 2001) over the survey volume for simplicity,
and therefore use the $z=0$ snapshot simulation data in two cosmological
models; Standard CDM (SCDM) with $\Omega_0=1$, $\lambda_0=0$, $h=0.5$,
$\Gamma=0.5$, and $\sigma_8=0.6$, and Lambda CDM (LCDM) with
$\Omega_0=0.3$, $\lambda_0=0.7$ $h=0.7$, $\Gamma=0.21$, and
$\sigma_8=1.0$, where $h$ is the Hubble constant in units of
100km/s/Mpc, $\Gamma$ is the shape parameter of the transfer function,
and $\sigma_8$ is the r.m.s. mass fluctuation amplitude at $8h^{-1}$Mpc.

To simulate the effect of the shape of the survey volume, we extract
``Wedge(r)'' samples of dark matter particles from the full simulation
cube data in real space so that they have exactly the same volume shape
as the Northern and Southern stripe regions.  To construct mock samples
that extend beyond the simulation box size, we duplicate particles using
the periodic boundary condition.  The unobserved gap regions are also
taken into account.  To examine the effect of redshift space
distortions, we construct ``Wedge(z)'' samples similar to the Wedge(r)
samples but in redshift space using the radial peculiar velocity of the
simulation particles to perturb the positions along the line of sight.

To include the selection function due to the apparent-magnitude limits,
we construct ``Wedge$(\phi)$'' samples from the Wedge(r) samples by
applying the radial selection function expected for our limits $14.5<r^*
<17.6$ and then add the peculiar velocity to each particle, so that the
resulting particle distribution matches the redshift distribution of the
observed galaxies (see subsection \ref{subsec:selection}).  Note that
this procedure implicitly assumes that the luminosity functions
constructed for the observed redshifts of objects do not suffer from the
redshift-distortion effect.  In order to check this, we also constructed
the corresponding ``Wedge$(\phi)$'' samples by directly applying the
selection function to the Wedge(z) samples.  Reassuringly, the two mock
samples constructed from the slightly different procedures turned out to
yield almost indistinguishable genus curves. This may be explained by
the fact that the luminosity functions are constructed with respect to
the mean redshift of objects, and thus the peculiar velocity effect is
statistically canceled.  Finally, we construct
``Wedge$(\phi\cdot\phi^{-1})$'' samples that are spatially the same as
the Wedge$(\phi)$ samples, but where each simulation particle is
weighted by the inverse of the selection function in the analysis so as
to correct for the selection function in a straightforward manner.  This
yields the same correction to the simulated density field that is
applied to the SDSS EDR galaxies when we compute the genus.

\begin{table}[h]
  \caption{Mock samples and SDSS EDR data adopted in the present
 analysis (for $z_{\rm max}=0.1$).}
  \label{tab:data}
 \begin{center}
 \begin{tabular}{c|c|c|c|c|c}
   Name & volume shape & real/redshift & $\phi(z)$ & $\phi^{-1}$
   correction & No. of particles/galaxies\\ \hline
   Full simulation  
& $(300h^{-1}{\rm Mpc})^3$ cube & real-space & No &  --- & $256^3$ \\
   Wedge (r) & wedge($z<z_{\rm max}$) & real-space & No & --- 
& $\sim 37000 (z_{\rm max}=0.1)$ \\
   Wedge (z) & wedge($z<z_{\rm max}$) & redshift-space & No & --- 
& $\sim 37000 (z_{\rm max}=0.1)$ \\
   Wedge $(\phi)$ & wedge($z<z_{\rm max}$) & redshift-space & Yes & No 
& $\sim 7500 (z_{\rm max}=0.1)$ \\
   Wedge $(\phi\cdot\phi^{-1})$ & wedge($z<z_{\rm max}$) &
  redshift-space & Yes & Yes & $\sim 7500 (z_{\rm max}=0.1)$ \\ 
   SDSS EDR (North) & wedge($z<z_{\rm max}$) & redshift-space & Yes & Yes 
& $7758 (z_{\rm max}=0.1)$ \\
   SDSS EDR (South) & wedge($z<z_{\rm max}$) & redshift-space & Yes & Yes 
& $6580 (z_{\rm max}=0.1)$ 
    \end{tabular}
  \end{center}
\end{table}

\section{The Genus Statistic}

\subsection{Computing the genus}

The genus, $G(\nus)$, is defined as $-1/2$ times the Euler
characteristic of the isodensity contour of the density field $\delta$
at the threshold level of $\nus$ times the r.m.s. fluctuation $\sigma$.
In practice this is equal to (number of holes) $-$ (number of isolated
regions) of the isodensity surface.  In this paper, we use $g(\nus)$ to
refer to the genus per unit volume, i.e., $g(\nus) \equiv G(\nus)/V$
where $V$ is the volume of the survey.  The genus $G$ of the isodensity
contour contained within the survey volume is obtained by integrating
the curvature of that surface using the Gauss-Bonnet theorem
\begin{equation}
G=-{1\over 4\pi} \int {1\over R_1 R_2} dA
\end{equation}
where $R_1$ and $R_2$ are the principal radii of curvature of the
surface.  This form of the genus allows for partial contributions to the
genus from structures that extend beyond the survey boundaries.

In this paper we primarily examine the genus as a function of the
threshold level $\nus$, which indicates that the isodensity surface is
drawn at $\nus$ times the r.m.s. density fluctuation $\sigma$ of the
density field after smoothing, as described below. Note carefully that
the description of the density threshold in terms of $\nus$ differs from
the definition of $\nu_{\rm f}$ used in some papers on the genus
statistics (e.g., Gott et al. 1989), in which $\nu_{\rm f}$ is used to
parameterize the fraction $f$ of the volume that lies on the high-density
side of the contour,
\begin{equation}
f = {1 \over \sqrt{2\pi}} \int_{\nuf}^\infty e^{-x^2/2} dx .
\label{eq:nuf}
\end{equation}
We label this volume-fraction definition of the threshold $\nuf$ in the
discussion below. In section 6 we present our final results in
terms of both $\nus$ and $\nuf$.

For a Gaussian random density field, the genus per unit volume is
\begin{equation}
\label{eq:genus_rd}
g_\RG(\nus) = \frac{1}{(2\pi)^2}\left(\frac{\sigma_1^2}{3\sigma^2}
\right)^{3/2}(1-\nus^2)\exp\left(-\frac{\nus^2}{2}\right) , 
\end{equation}
where $\sigma_1 \equiv \langle|\nabla\delta|^2\rangle^{1/2}$ and $\sigma
\equiv \langle\delta^2\rangle^{1/2}$.  Here $\nabla\delta$ denotes the
spatial derivative of the density field $\delta$ and
$\langle\cdot\cdot\cdot\rangle$ is the average over the
probability distribution function (PDF) of $\delta$ and $\nabla\delta$
(c.f., \cite{D1970, A1981, BBKS1986, GMD1986, HGW1986}).

\begin{figure}[htb]
\begin{center}
\FigureFile(45mm,45mm){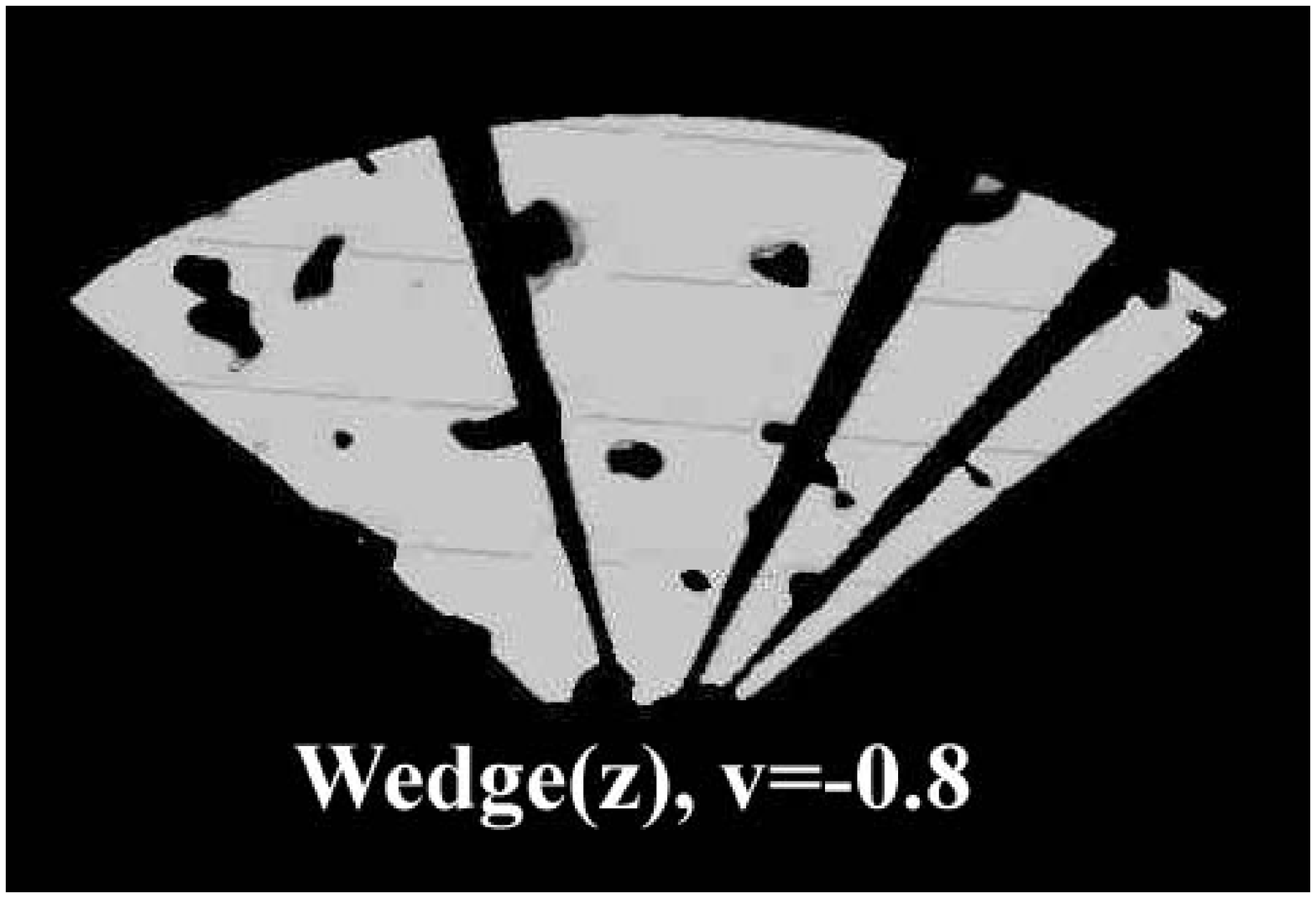}
\FigureFile(45mm,45mm){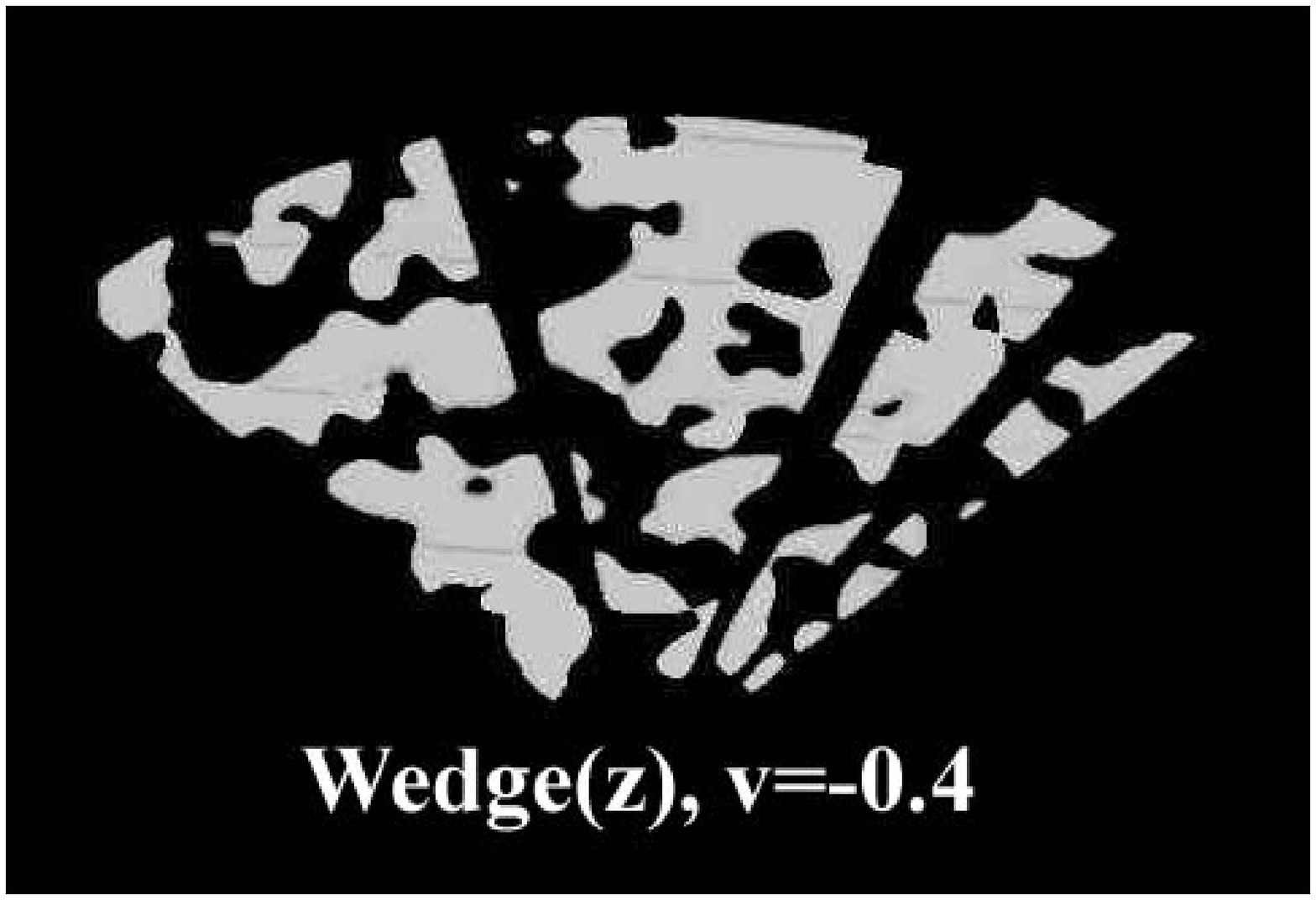}
\FigureFile(45mm,45mm){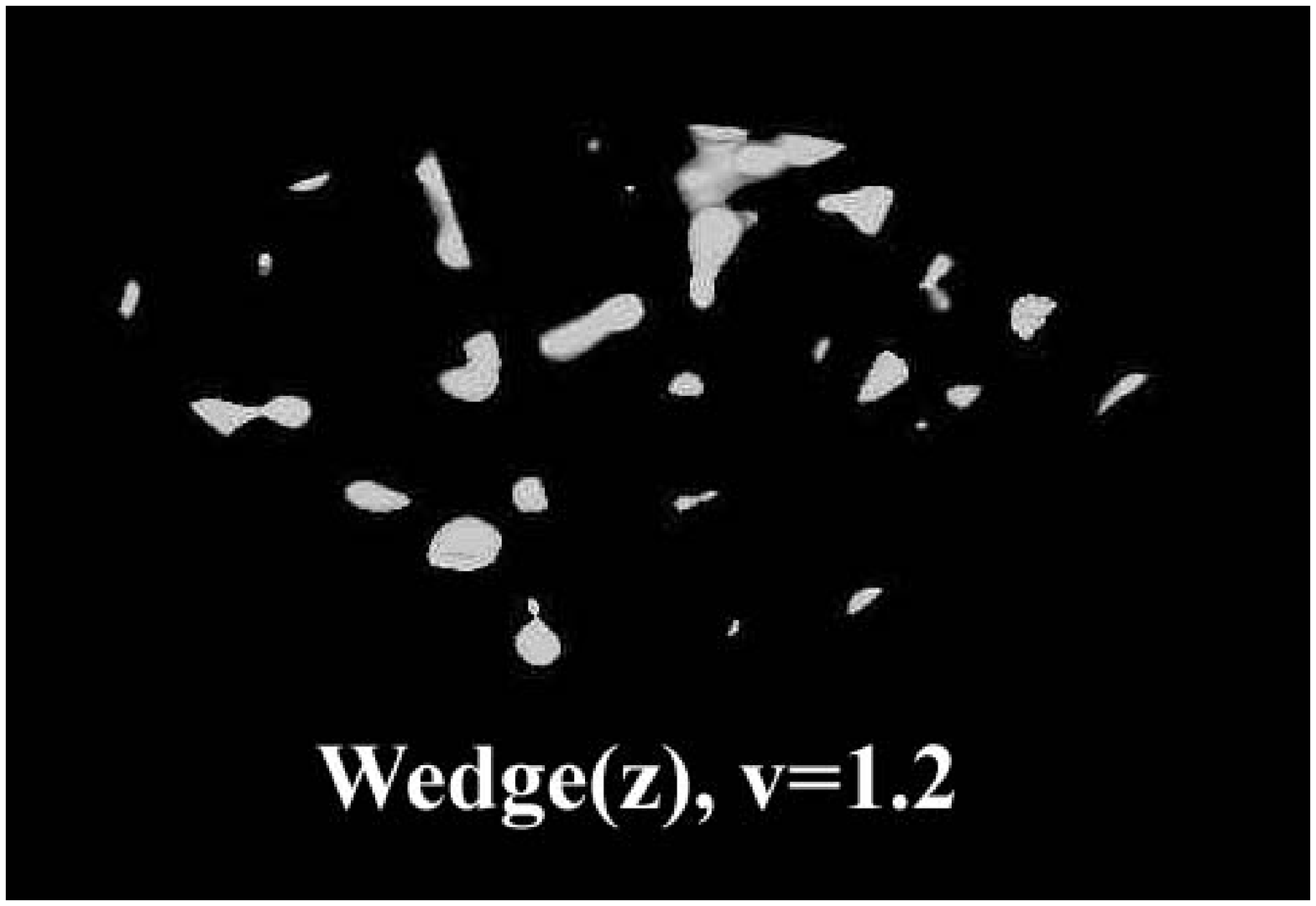}
\FigureFile(45mm,45mm){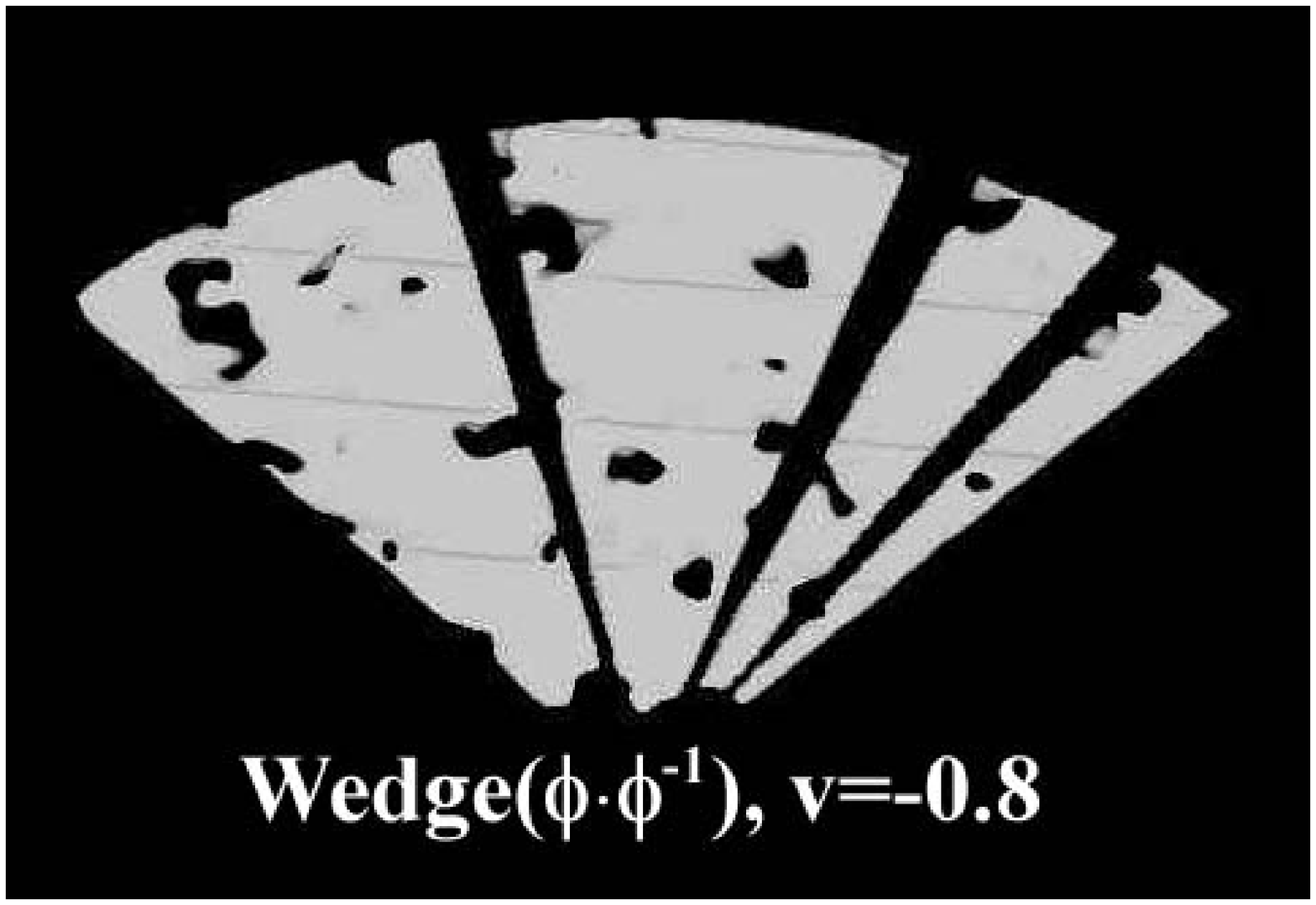}
\FigureFile(45mm,45mm){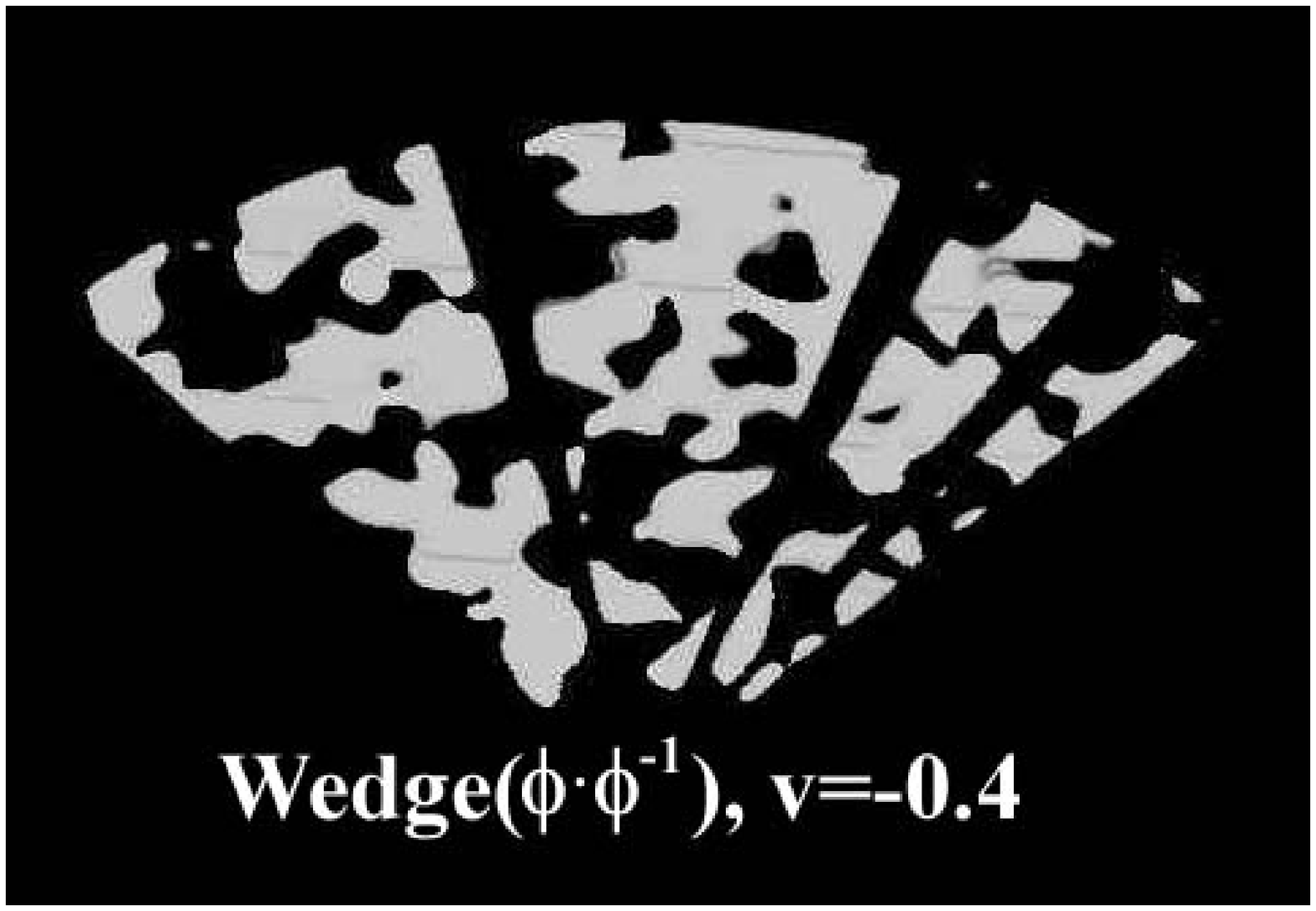}
\FigureFile(45mm,45mm){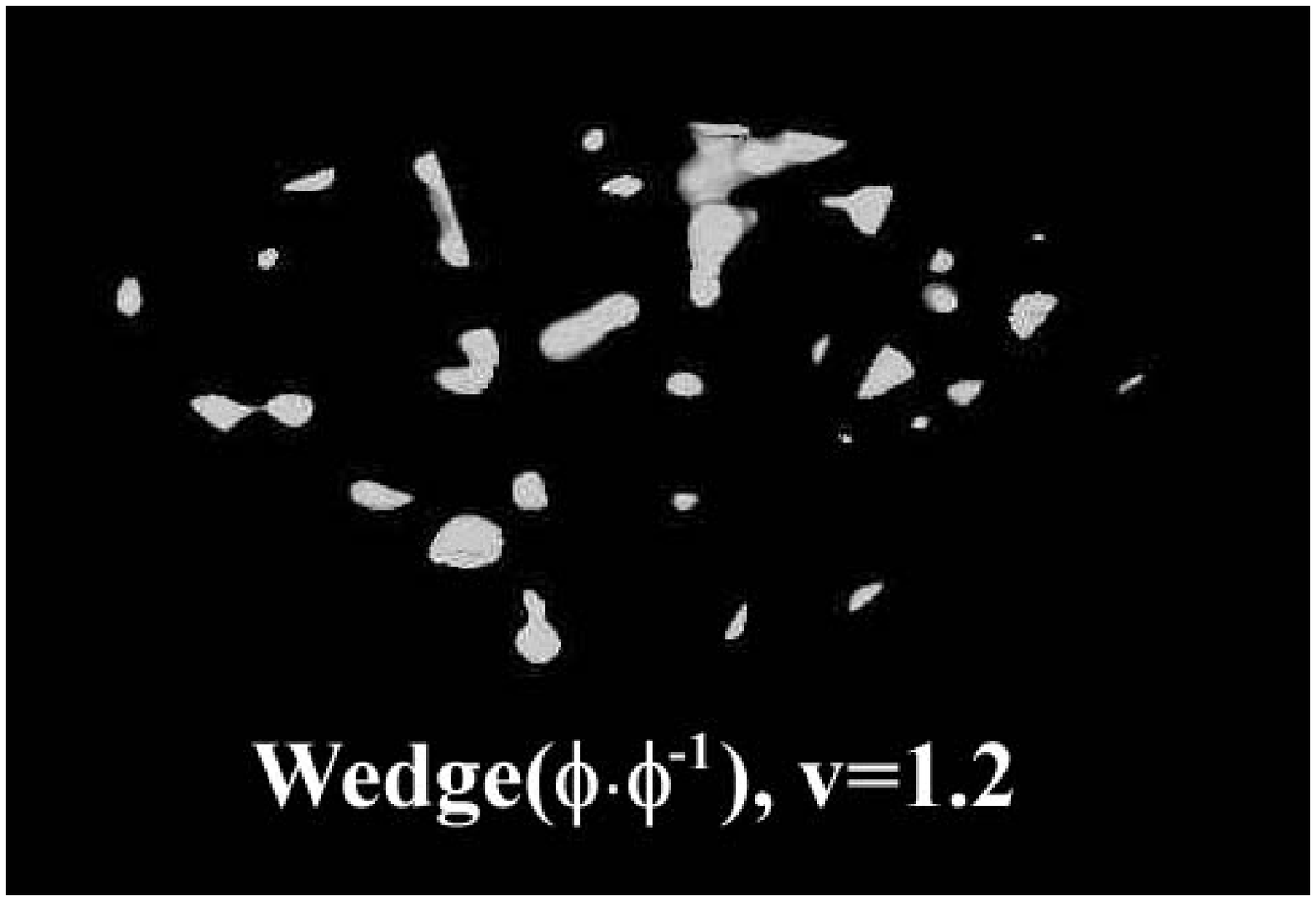}
\FigureFile(45mm,45mm){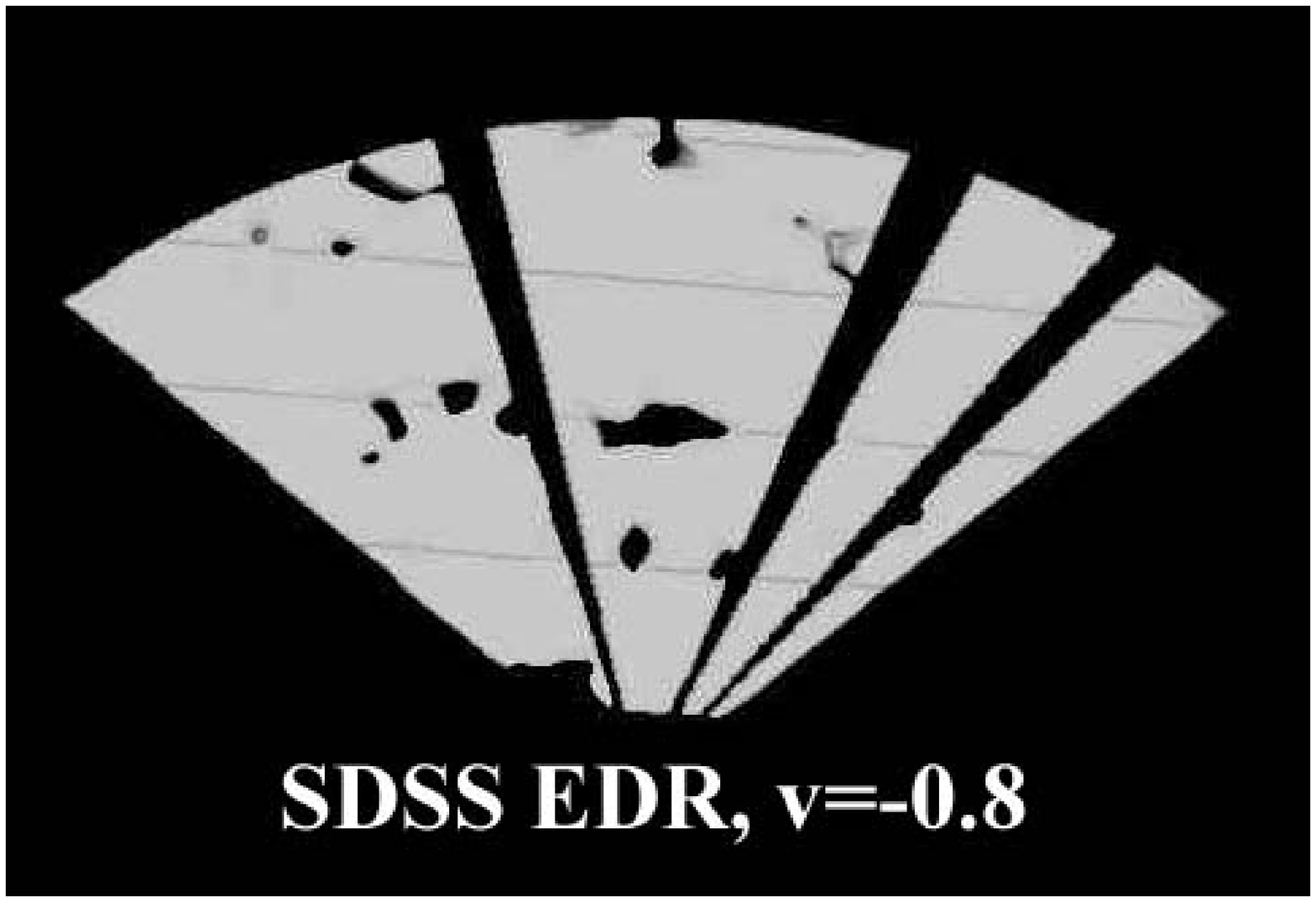}
\FigureFile(45mm,45mm){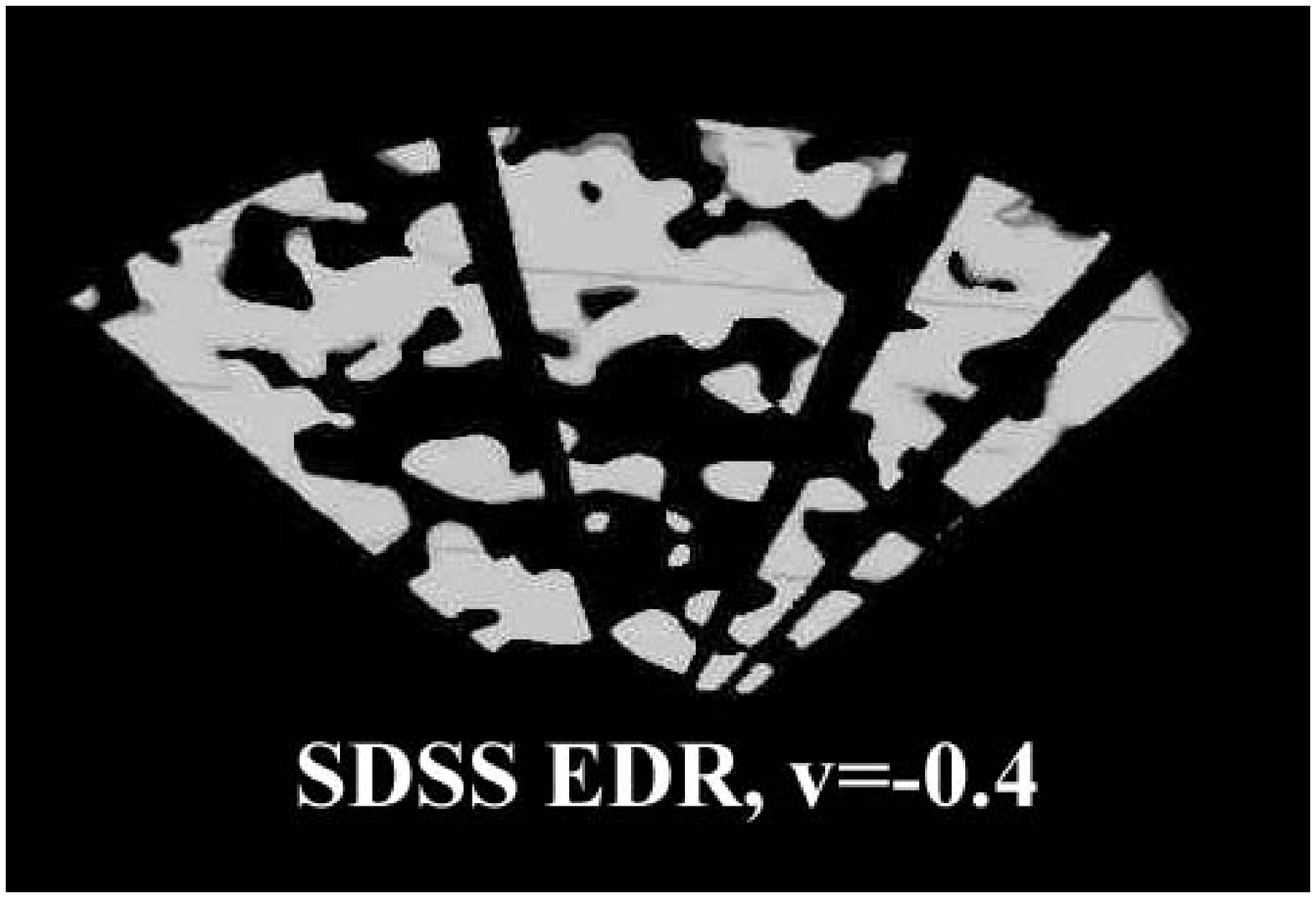}
\FigureFile(45mm,45mm){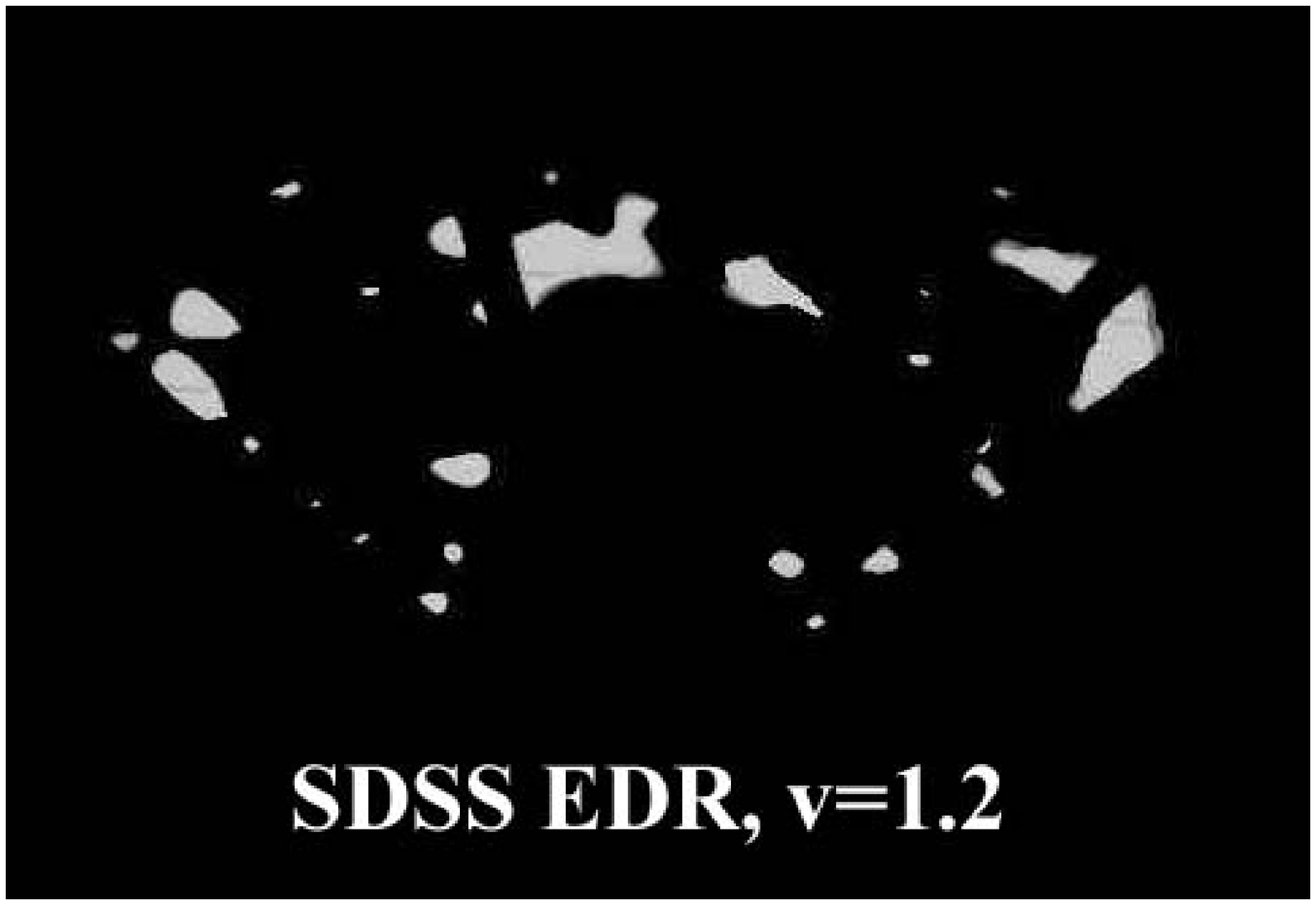}
\caption{Isodensity contour surfaces for different simulation mock
  samples and SDSS EDR galaxies. The Gaussian smoothing length is
  $R_{\rm G}=5h^{-1}$Mpc .  From top to bottom are contours for Wedge(z),
  Wedge($\phi\cdot\phi^{-1}$), and SDSS EDR galaxies (corrected for
  the selection function).  Left, center and right panels illustrate
  the isodensity contours corresponding to $\nus = -0.8$, $-0.4$ and $1.2$.
  }
\label{fig:isocontour}
\end{center}
\end{figure}

We compute the genus for density fields from mock samples and SDSS
galaxies using the CONTOUR 3D routine \citep{W1988}.  We use the
cloud-in-cell (CIC) method to assign survey galaxies and dark matter
particles to cells on a $256^3$ grid.  We Fourier transform the density
field, multiply it by the Fourier transform of a Gaussian window with
smoothing length $R_{\rm G}$, then transform it back to real space.  We
use the conventional definition of the Gaussian smoothing length $R_{\rm
G}$ for the smoothing kernel
\begin{equation}
W_{\rm G}(r)=\frac{1}{\sqrt{2\pi}R_{\rm G}}
\exp\left(-\frac{r^2}{2R_{\rm G}^2}\right) .
\end{equation}
This definition of $R_{\rm G}$ differs from some previous papers
in which the ``smoothing length'' is defined as
$\lambda_{\rm G} = \sqrt{2} R_{\rm G}$.  The smoothed density fields
are used to define the isodensity surfaces with a given threshold
$\nus$. The genus is evaluated by integrating the deficit angles
(contributions to the curvature at grid cell vertices) over the
isodensity surfaces. In the wedge mock samples, integration of the
deficit angles is performed only at grid cells located inside the
wedge region; the contribution at grid cells on the boundary and
outside the region is neglected.  Examples of isodensity surfaces for
the different mock samples as well as for the SDSS galaxies are
illustrated in Figure \ref{fig:isocontour}.

\subsection{Log-normal model prediction vs. Full simulation}

Since the one-point PDF of dark matter particles in our simulations is
well approximated by the log-normal PDF (Kayo et al. 2001), it is
natural to expect that the log-normal model can be used to predict the
genus.  \citet{MY1996} derived the genus expression assuming that the
nonlinear density field of dark matter has a one-to-one mapping to its
primordial Gaussian field. If one adopts the log-normal mapping, the
result is explicitly written as
\begin{eqnarray}
\label{eq:genus_dm}
g_{\rm mass,\LN}(\num) & = & g_{\rm max,\LN}[1-x_\LN^2(\num)]
\,\exp \left[-\frac{x_\LN^2(\num)}{2}\right] , \\
x_\LN(\num) &\equiv& \frac{\ln[(1+\num\sigmamm)\sqrt{1+\sigmamm^2}]}
{\sqrt{\ln(1+\sigmamm^2)}} , 
\end{eqnarray}
where $\num=\deltam/\sigmamm$, $\sigmamm=\langle
\deltam^2\rangle^{1/2}$, $\deltam$ is the mass density fluctuation,
and the maximum value, $g_{\rm max,\LN}$ is given by:
\begin{equation}
\label{eq:genus_max_ln}
g_{\rm max,\LN}  = \frac{1}{(2\pi)^2}
\left[\frac{\langle |\nabla\deltam |^2\rangle}
{3(1+\sigmamm^2)\log(1+\sigmamm^2)} \right]^{3/2} .
\end{equation}
The one-to-one mapping assumption is much stronger than the statement
that the one-point PDF is empirically fitted to the log-normal PDF. In
particular, the prediction (eq. [\ref{eq:genus_dm}]) accurately
describes previous simulation results for dark matter particles
(Matsubara \& Yokoyama 1996; Hikage, Taruya \& Suto 2002).

\begin{figure}[thp]
\begin{center}
\FigureFile(100mm,100mm){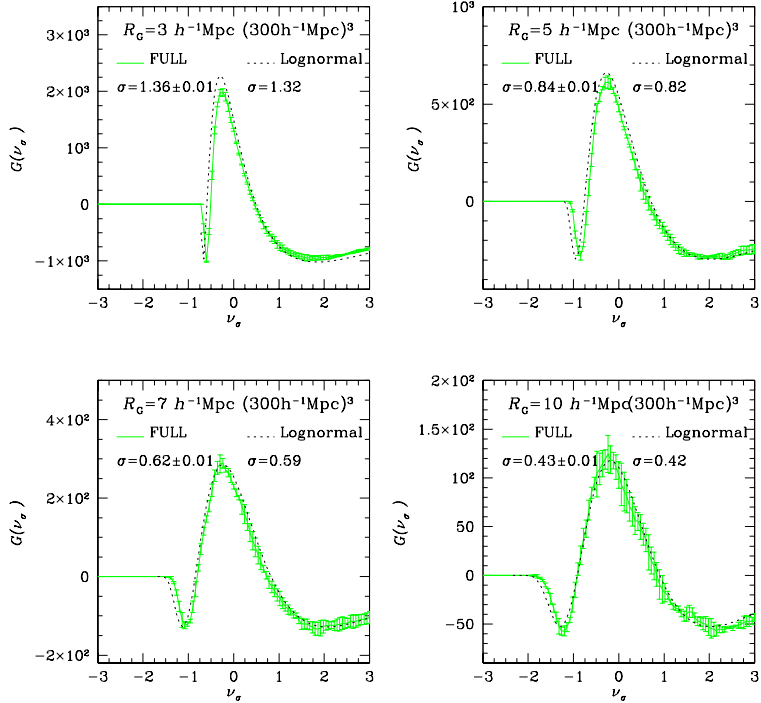}
\end{center}
\caption{Genus $G(\nus)$ 
  as a function of $\nus \equiv \delta/\sigma$ for the particle
  distribution in the full N-body simulation data (solid lines)
  compared with the log-normal model predictions (dotted lines) for
  several Gaussian smoothing lengths $R_{\rm G}$.  The values of
  $\sigma$ in each panel show the rms density fluctuations on the
  corresponding smoothing scale.}
\label{fig:full_lognormal}
\end{figure}

Figure \ref{fig:full_lognormal} compares the total genus, $G(\nus)$, for
the particle distribution in the full $(300h^{-1}{\rm Mpc})^3$ cubic
volume with the log-normal prediction (eq.[\ref{eq:genus_dm}]).  The
plotted error bars represent the standard deviation among three
different realizations of full simulations.  We use Gaussian smoothing
lengths $R_{\rm G}=3h^{-1}$Mpc (Upper-left), $5h^{-1}$Mpc (Upper-right),
$7h^{-1}$Mpc (Lower-left), and $10h^{-1}$Mpc (Lower-right).  Also shown
are the r.m.s. mass fluctuation amplitudes, $\sigma=\sigma(R_{\rm G})$
on the Gaussian smoothing scale $R_{\rm G}$.  The values of $\sigma$ and
the quoted standard deviation for simulations are directly evaluated
from the three different realizations, while those for the log-normal
model prediction are {\it not} fitted to the simulation data, but rather
are computed using the nonlinear power spectrum of Peacock \& Dodds
(1996) and the assumed set of cosmological parameters .  In this sense,
this comparison does not have any free fitting parameters.  Figure
\ref{fig:full_lognormal} shows that the log-normal model describes well
both the amplitude and the shape of $G(\nus)$ for the simulation data at
all smoothing lengths.

\section{Observational systematic effects on genus 
statistics evaluated with mock samples}

In what follows, we begin with the full simulation particle distribution
and consider the key observational effects in a cumulative manner: the
survey volume geometry, redshift-space distortion, radial selection
function due to the magnitude-limit, and correction for the selection
function. These effects are discussed in the following plots and
subsections.  In this section, we show results of mock samples that
mimic the Northern stripe using simulations of the LCDM model.

\subsection{Full Simulation vs. Wedge(r): Effect of survey shape}

\begin{figure}[thp]
\begin{center}
\FigureFile(100mm,100mm){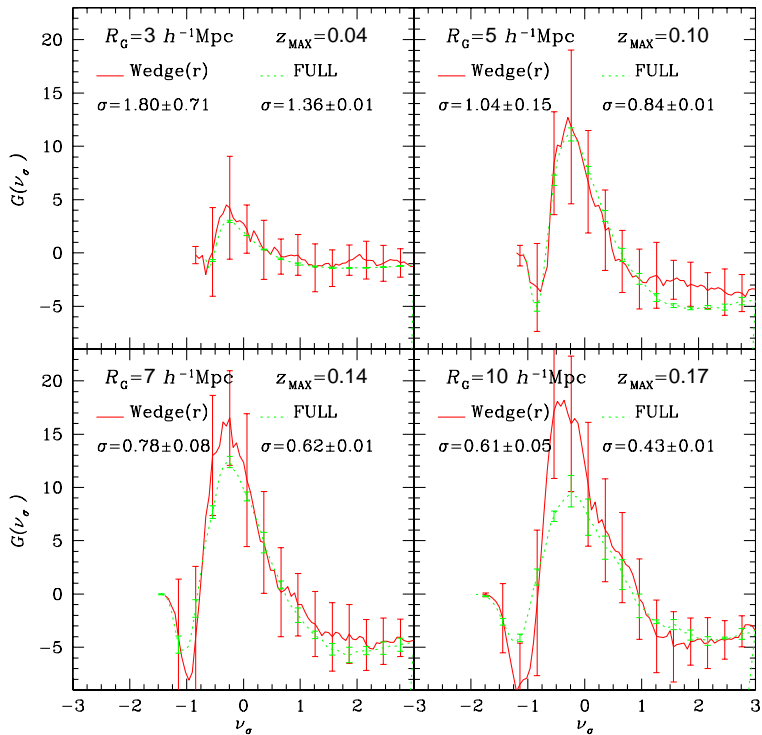}
\end{center}
\caption{Effect of the survey volume shape on $G(\nus)$ with different
  $R_{\rm G}$.  Solid lines show results for the full simulations.
  Dotted lines show the genus for the Wedge(r) mock samples. The plotted
  error-bars represent the one standard deviation computed from the 15
  independent Wedge(r) mock samples sampled at $\Delta\nus=0.25$.  To
  make this comparison, the genus amplitude of the full simulation is
  normalized by the ratio of the full and wedge volumes.}
  \label{fig:full_wedge}
\end{figure}

Figure \ref{fig:full_wedge} shows the effect of the survey volume
shape by comparing the two mock samples, Full and Wedge(r).  The
volume of Wedge(r) is selected to reproduce the geometry
of SDSS EDR North. The mean and standard deviation of the genus
are estimated from fifteen independent wedge samples drawn from the
three different realizations of the full simulations. To account for
the survey geometry, the CONTOUR3D routine uses reference mask
particles randomly distributed over the survey region. The density
field of mask particles is computed on a grid with the CIC method,
then smoothed in the same way as the dark matter particles evolved in
the simulations (e.g., Moore et al. 1992, Melott \& Dominik 1993).

In an apparent-magnitude limited galaxy survey, the number density of
observed galaxies decreases with redshift.  Reliable reconstruction of
density fields from the discrete particle distribution requires that the
smoothing length $R_{\rm G}$ approximately exceeds the mean separation
length of galaxies at the redshift.  (Using too small a smoothing length
would result in isodensity contours that are spheres around each
particle.)  To satisfy this criterion, we set the upper limit of the
redshift $z_{\rm max} = z_{\rm max}(R_{\rm G})$ so that $R_G$ equals the
mean separation of the magnitude-limited galaxy sample at $z_{\rm max}$
(Vogeley et al. 1994).  The values of $z_{\rm max}$, the resulting
numbers of galaxies and the minimum values of the absolute magnitude
$M_{\rm min}$ for SDSS EDR galaxy subsamples that we consider below are
listed in Table \ref{tab:R_zmax} as a function of $R_{\rm G}$. We also
list the number of independent resolution elements in each sample,
\begin{equation}
N_{\rm res} \equiv 
\frac{V_{\rm survey}(<z_{\rm max})}{(2\pi)^{3/2} R_{\rm G}^3} ,
\end{equation}
which provides an approximate measure of the sample variance in the
current analysis, roughly proportional to $N_{\rm res}^{1/2}$. Gott et
al. (1989) noted that $N_{\rm res} \approx 100$ yields a fractional
error of $\sim 25\%$ in estimating the genus curve.  Comparison of $N_{\rm
res}$ as a function of $R_{\rm G}$ shows that our results for
$R_{\rm G}=5h^{-1}$Mpc and $R_{\rm G}=7h^{-1}$Mpc are statistically more
reliable than those for the other two smoothing scales.

\begin{table}[thb]
  \caption{Samples of SDSS EDR galaxies analyzed for different smoothing
lengths $R_{\rm G}$ assuming LCDM model parameters.} \label{tab:R_zmax}
 \begin{center}
 \begin{tabular}{c|c|c|c|c|c|c|c}
 & & \multicolumn{3}{c|}{Northern stripe} 
    & \multicolumn{3}{c}{Southern stripe} \\ \cline{3-8}
 \raisebox{1.5ex}[0pt]{$R_{\rm G}$} & \raisebox{1.5ex}[0pt]{$z_{\rm max}$}
 & number of galaxies & $N_{\rm res}$ & $M_{\rm min}$ &  
 number of galaxies & $N_{\rm res}$ & $M_{\rm min}$ \\ \hline
   3 & 0.04 & 1275 & 94 & $-21.2$ & 1730 & 90 & $-21.1$ \\ 
   5 & 0.10  & 7758 & 248 & $-22.8$ & 6580 & 236 & $-22.6$ \\ 
   7 & 0.14  & 11294 & 214 & $-23.1$ & 9050 & 203 & $-22.9$ \\ 
   10 & 0.17  & 12770 & 132 & $-23.1$ & 10203 & 126 & $-23.2$
    \end{tabular}
  \end{center}
\end{table}

To correct for the different volumes of the full and wedge samples, we
divide the genus of the full sample by the appropriate volume ratio.
Comparison of the genus shows that the Wedge(r) mock samples yield a
larger genus amplitude than that of the full simulations, especially for
$R_{\rm G} > 7h^{-1}$Mpc.  Because the thickness of the wedge survey
region of SDSS EDR is merely $13h^{-1}$ Mpc at $z=0.1$, a large fraction
of the smoothing region lies outside the survey boundary, which leads to
the unphysical behavior revealed here.  In other words, our 3D analysis
essentially reduces to a 2D analysis for $R_{\rm G} > 7h^{-1}$Mpc. Some
of the clustering signal is effectively smoothed to beyond the survey
boundary and not taken into account in the analysis which only includes
grid vertices within the survey volume. Thus the signal-to-noise ratio
for structure within the survey is somewhat reduced.  Therefore it may
not be meaningful to attempt to understand the detailed behavior of the
genus curve on these larger smoothing scales for such a geometry.
Nevertheless one can still explore the cosmological implications by
comparing the results between the (final) mock samples and the observed
galaxies, since both data and simulations are treated in the same
fashion.

\subsection{Wedge(z) vs. Wedge$(\phi\cdot\phi^{-1})$: 
radial selection effect
\label{subsec:selection}}

The Wedge(z) samples differ from the Wedge(r) samples only by
introduction of the distortion of redshift space caused by peculiar
velocities.  This distortion has minimal effect on the genus, consistent
with previous results (Matsubara 1996; Matsubara \& Suto 1996). Thus,
the more important observational effect comes from the
apparent-magnitude limit of the sample, which changes the mean galaxy
number density with redshift.  Appropriate correction for this selection
function is necessary to accurately define the isodensity surfaces in
apparent-magnitude limited samples.

\begin{figure}[htp]
\begin{center}
\FigureFile(100mm,100mm){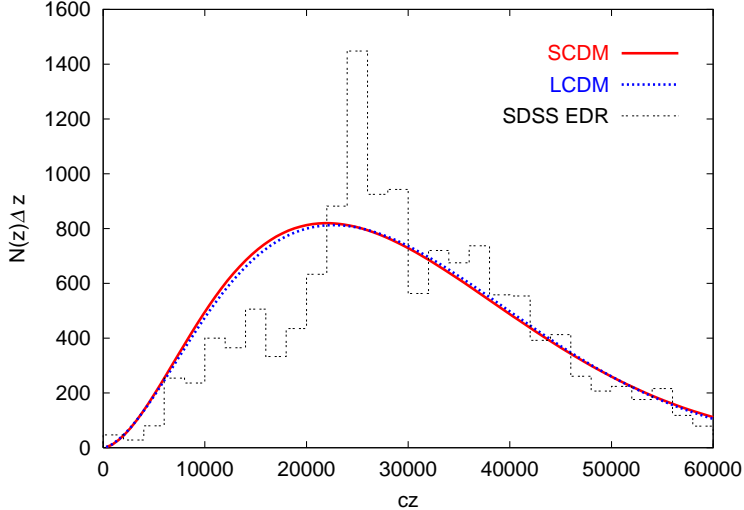}
\end{center}
\caption{Number density distribution of SDSS EDR galaxies. The redshift
width $\Delta z$ of the histogram is $1/150$. Fits from equation
(\ref{eq:phiz}) are plotted assuming the cosmological parameters of the
LCDM (solid line) and SCDM (dotted line) models.  } \label{fig:Nz}
\end{figure}

We compute the selection function, $\phi(z)$, by integrating over the
Schechter form of the luminosity function,
\begin{equation}
\label{eq:phiz}
\phi(z)
\propto \int^{M_{\rm max}(z)}_{M_{\rm min}(z)}
\phi_\ast[10^{0.4(M_\ast-M)}]^{\alpha+1}
\exp[-10^{0.4(M_\ast-M)}]dM ,
\end{equation}
with parameters measured in the $r'$ band by Blanton et al. (2001) for
the SDSS, $M_\ast-5\log_{10} h=-20.83 (-20.67)$, $\alpha=-1.20 (-1.15)$,
$\phi_\ast=1.46 (1.87)\times 10^{-2} h^3{\rm Mpc}^{-3}$ for analyses
that assume the cosmological parameters of the LCDM (SCDM) cosmological
model. The limits of integration are the absolute magnitude limits
\begin{equation}
M_{\rm max/min}(z)=m_{\rm max/min}
-5\log[(1+z)r(z_{\rm max/min})/10{\rm pc}]-K(z) .
\end{equation}
We use apparent-magnitude limits $m_{\rm max}=17.6$, $m_{\rm
  min}=14.5$, following Blanton et al. (2001), and apply an
approximate K-correction factor $K(z) = 0.9z$ valid for the typical
galaxy color of $g^\ast - r^\ast=0.65$ (Fukugita et al. 1995).  Figure
\ref{fig:Nz} shows the number density distribution of SDSS EDR
galaxies (histogram) compared with fits from equation (\ref{eq:phiz})
for the cosmological parameters of both the LCDM and SCDM models.

\begin{figure}[thp]
\begin{center}
\FigureFile(100mm,100mm){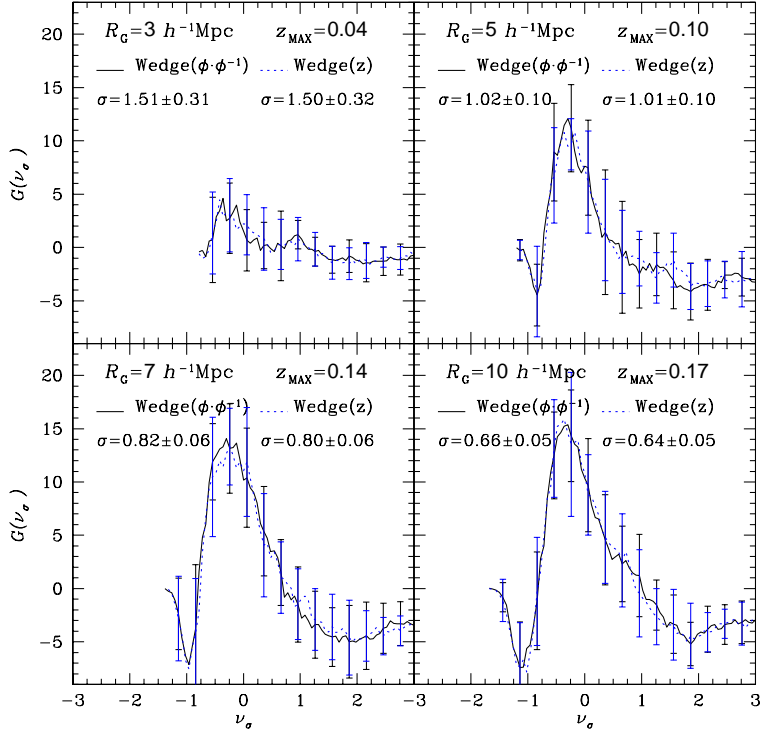}
\end{center}
\caption{Genus, $G(\nus)$, 
for Wedge(z) (solid lines) and Wedge$(\phi\cdot\phi^{-1})$ mock samples,
for different smoothing lengths $R_{\rm G}$.
This comparison shows that weighting galaxies in an apparent-magnitude
limited sample by the inverse of the selection function provides a
correct estimate of the genus.
}
\label{fig:correction}
\end{figure}

We randomly select dark matter particles in the Wedge(z) sample
according to the above selection function to reproduce the number of
observed SDSS galaxies in the current sample.  Then we attempt to
correct for the selection effect simply by weighting each particle by
the inverse of the selection function $\phi(z)^{-1}$ (e.g., Rhoads et
al. 1994; Vogeley et al. 1994) before smoothing the density field.  The
contours shown in Figure \ref{fig:isocontour} adopt this correction,
which illustrates that this correction accurately reproduces the
structure. Figure \ref{fig:correction} shows more quantitatively that
the correction is very accurate for the genus statistics. Therefore we
compare the genus of the SDSS galaxies and the mock samples after
applying the $1/\phi(z)$ correction to their redshift-space
distributions.

\section{Genus of the SDSS EDR Sample}

\begin{figure}[thp]
\begin{center}
\FigureFile(80mm,80mm){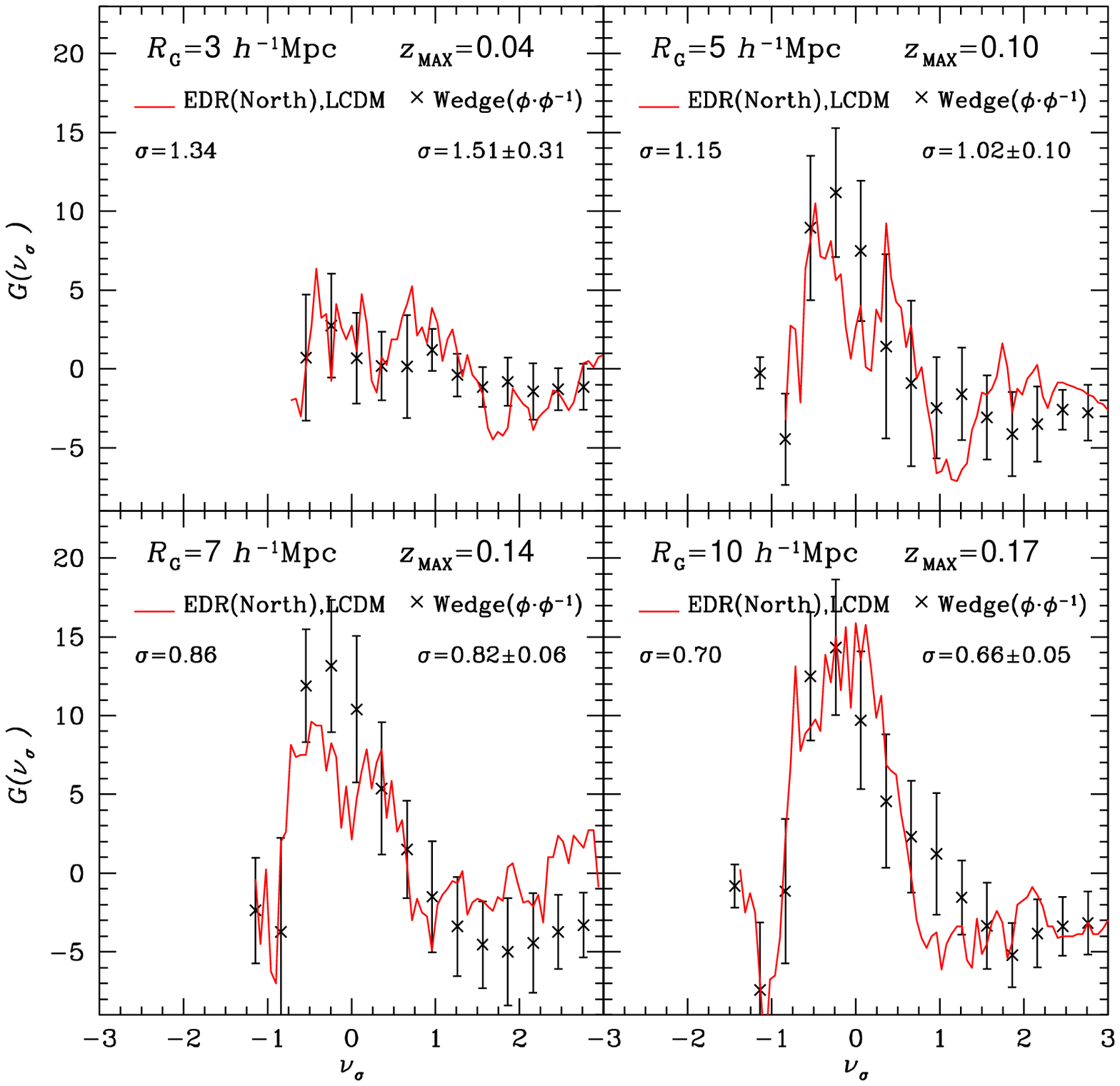}
\FigureFile(80mm,80mm){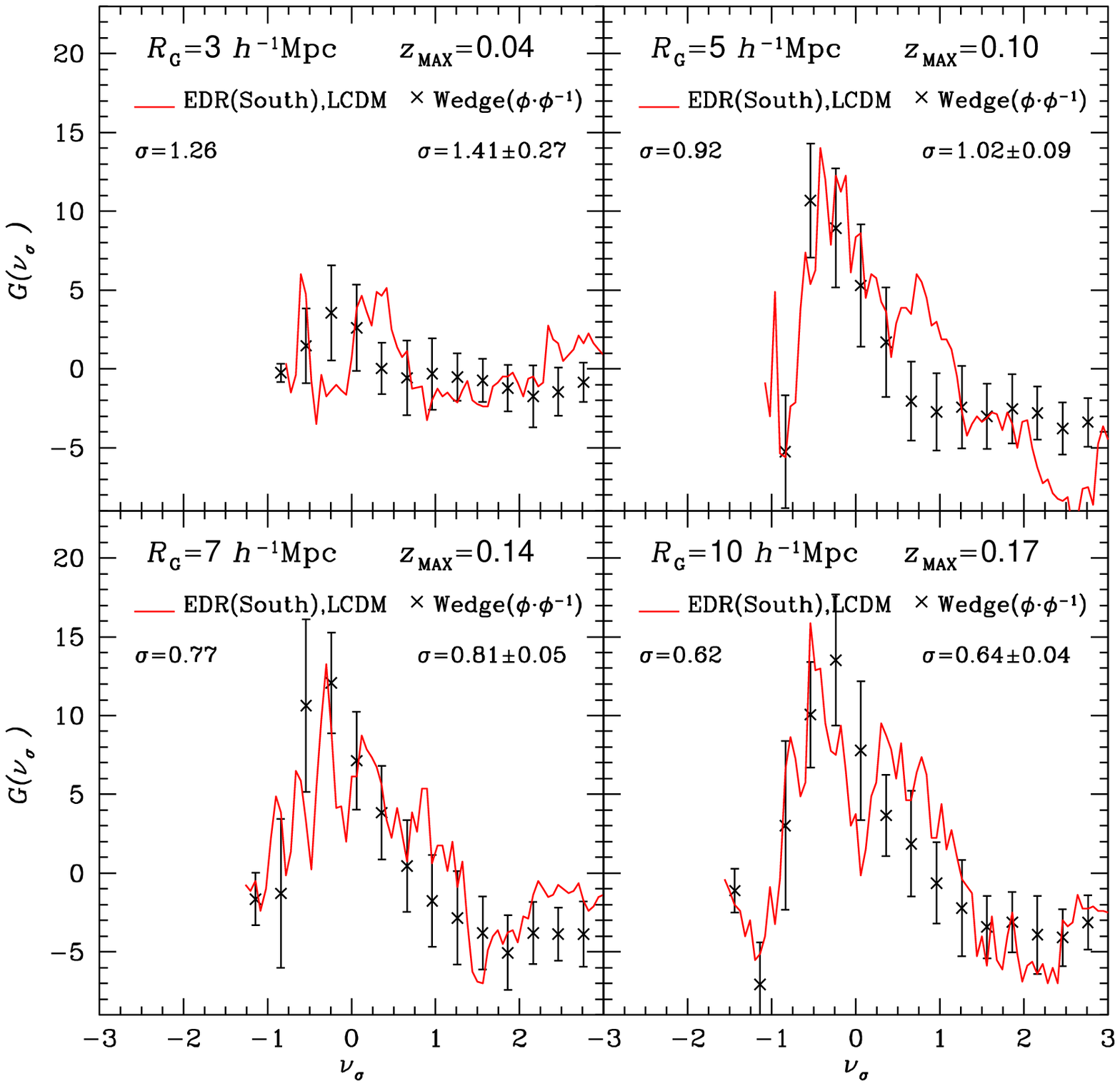}
\end{center}
\caption{Comparison of the genus $G(\nus)$ for SDSS EDR galaxies
  with Wedge$(\phi\cdot\phi^{-1})$ mock samples from LCDM simulations.
  Plotted separately are the genus curves for the Northern stripe
  ({\it Left}) and Southern stripe ({\it Right}) of the SDSS EDR
  sample. Error bars are shown for the LCDM prediction}  
\label{fig:sdss_mock_LCDM}
\end{figure}

\begin{figure}[thp]
\begin{center}
\FigureFile(80mm,80mm){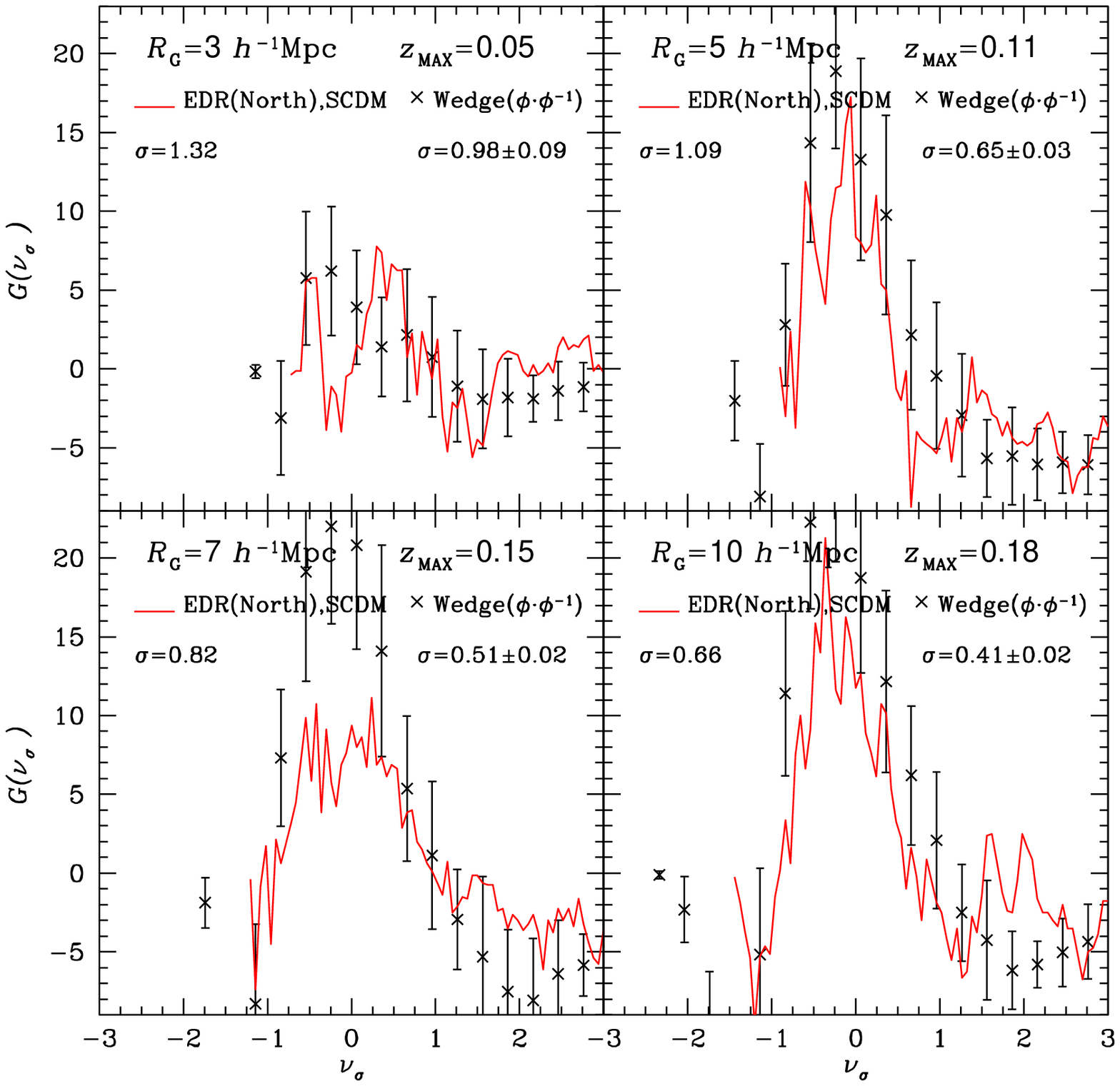}
\FigureFile(80mm,80mm){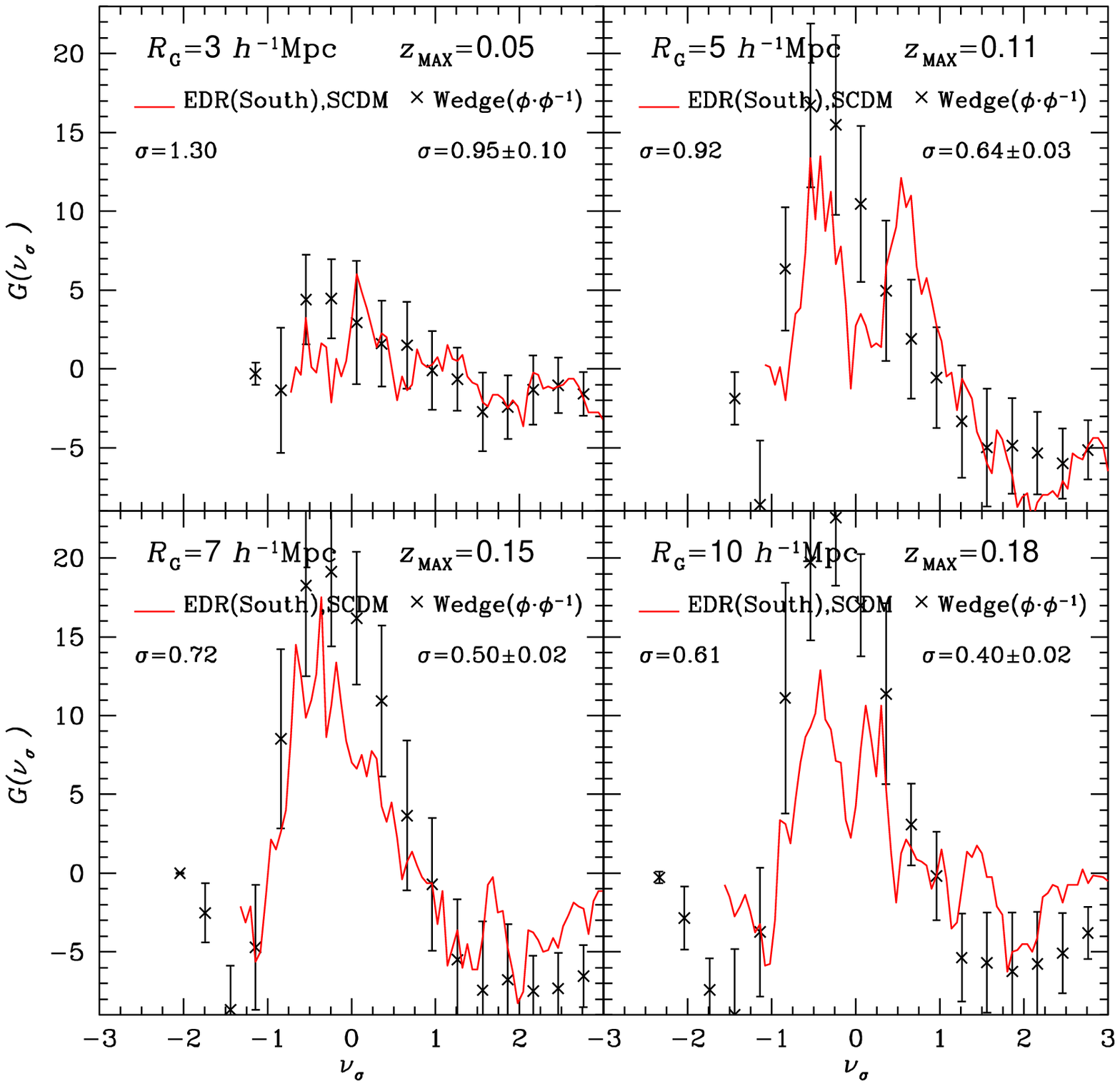}
\end{center}
\caption{Same as 
Figure \ref{fig:sdss_mock_LCDM} but in the
SCDM model.}  
\label{fig:sdss_mock_SCDM}
\end{figure}

Finally, we present the genus estimated for the apparent-magnitude
limited SDSS EDR galaxy samples. The galaxy density fields are
corrected for the selection function as discussed above.  Figures
\ref{fig:sdss_mock_LCDM} and \ref{fig:sdss_mock_SCDM} show the genus
curves ({\it filled circles}) for the SDSS galaxies assuming the
cosmological parameters of the LCDM and SCDM models, respectively, 
and the genus of the corresponding mock samples.

The plotted error-bars for each model prediction represent the 1
standard deviation computed from 15 independent mock samples.  To within
these uncertainties, we conclude that the genus for the SDSS EDR
galaxies is consistent with the LCDM model prediction, but does not
agree with that of SCDM.

Galaxy biasing is another source of uncertainty for relating the
observed genus curves to those obtained from the mock samples generated
from the distribution of dark matter particles (e.g., Colley et
al. 2000; Hikage, Taruya, Suto 2001). If LCDM is the correct
cosmological model, then the good match of the genus for mock samples
from the LCDM simulations to the observed SDSS genus may indicate that
nonlinearity in the galaxy biasing is relatively small, at least small
enough that it does not impact the genus statistic.  In separate work,
Kayo et al. (2002) showed that the biasing parameter for the SDSS EDR
galaxies is close to unity and that this biasing is approximately
scale-independent if LCDM is assumed. This linear biasing does not
affect the genus curve, thus our conclusion remains unchanged.


Differences between the genus curves for the mock samples drawn from
simulations of the LCDM and SCDM models arise largely from the power
spectra for those models which affect only the amplitude of the genus
curve (see eqs.[\ref{eq:genus_rd}] and [\ref{eq:genus_dm}] --
[\ref{eq:genus_max_ln}]).  The shapes of the genus curves for the two
models are almost identical.  In other words, the present genus analysis
places constraints on the cosmological models that are consistent with
previous results using the 2PCF, but it does not yet exploit the full
advantage of the genus statistics as a complementary statistic to the
conventional 2PCF. Nevertheless, it is encouraging that one can draw
meaningful cosmological conclusions from the genus analysis on the basis
of the available EDR sample from the genus amplitude alone.

Thus far in this paper we have plotted the genus curve as a function of
the density threshold parameterized by the number $\nus$ of standard
deviations above the mean of the smoothed density field.  In contrast,
the genus is often plotted as a function of the threshold $\nuf$ which
is defined so that the volume fraction on the high-density side of the
isodensity surface is identical to the volume in regions with density
contrast $\delta =\nuf \sigma$, for a Gaussian random field with
r.m.s. density fluctuations $\sigma$ (see Equation \ref{eq:nuf}).  If
the evolved density field has a one-to-one correspondence with the
initial random-Gaussian field, then this transformation removes the
effect of evolution of the PDF of the density field (see section 4.2).
Under this assumption, the genus as a function of volume fraction,
expressed as $G(\nuf)$, is sensitive only to the topology of the
isodensity contours rather than evolution with time of the density
threshold assigned to a contour.  Limitations of the approximation of
monotonicity in the relation between initial and evolved density fields
are examined by Kayo et al.  (2001).

\begin{figure}[thp]
\begin{center}
\FigureFile(80mm,80mm){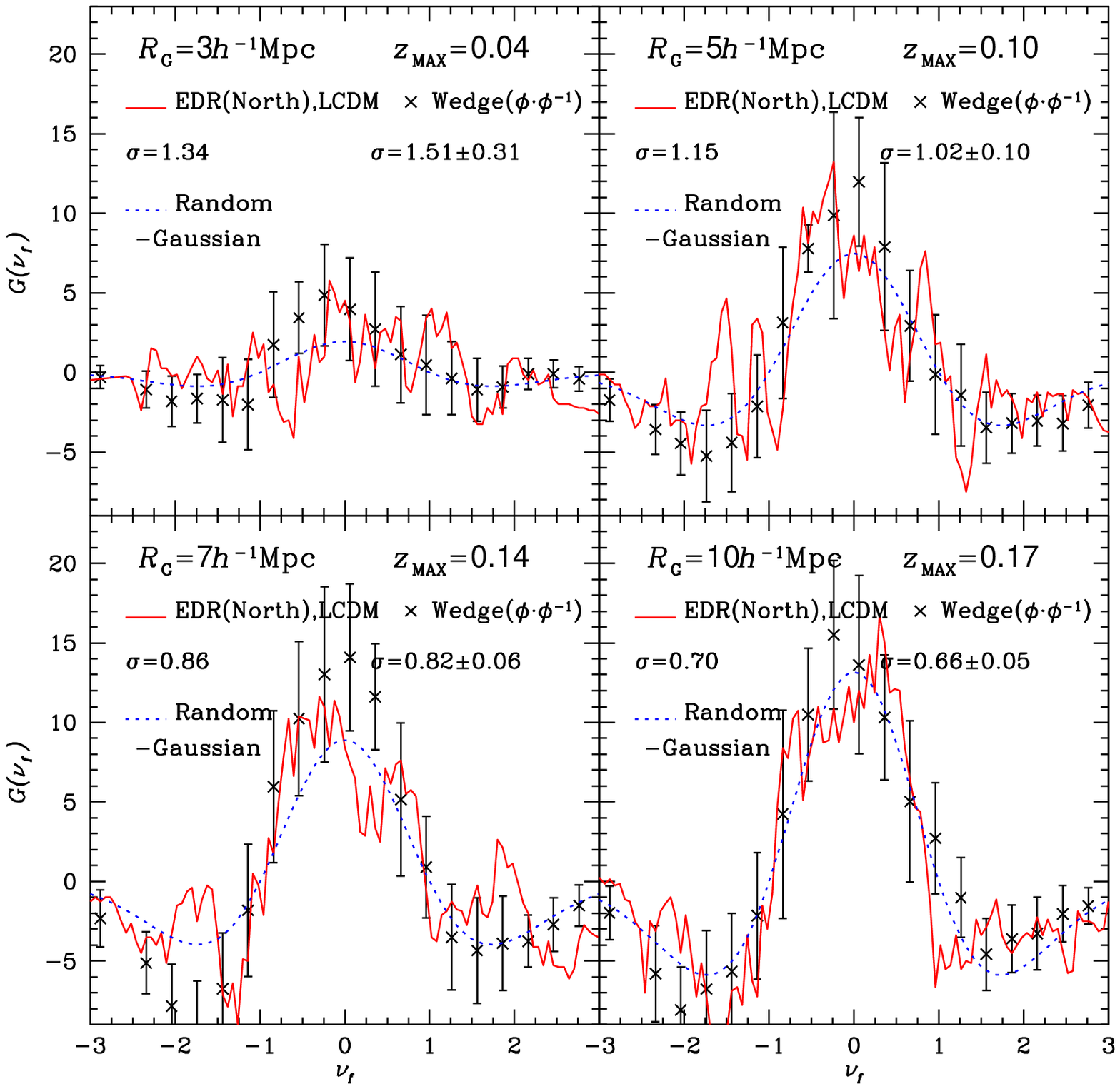}
\FigureFile(80mm,80mm){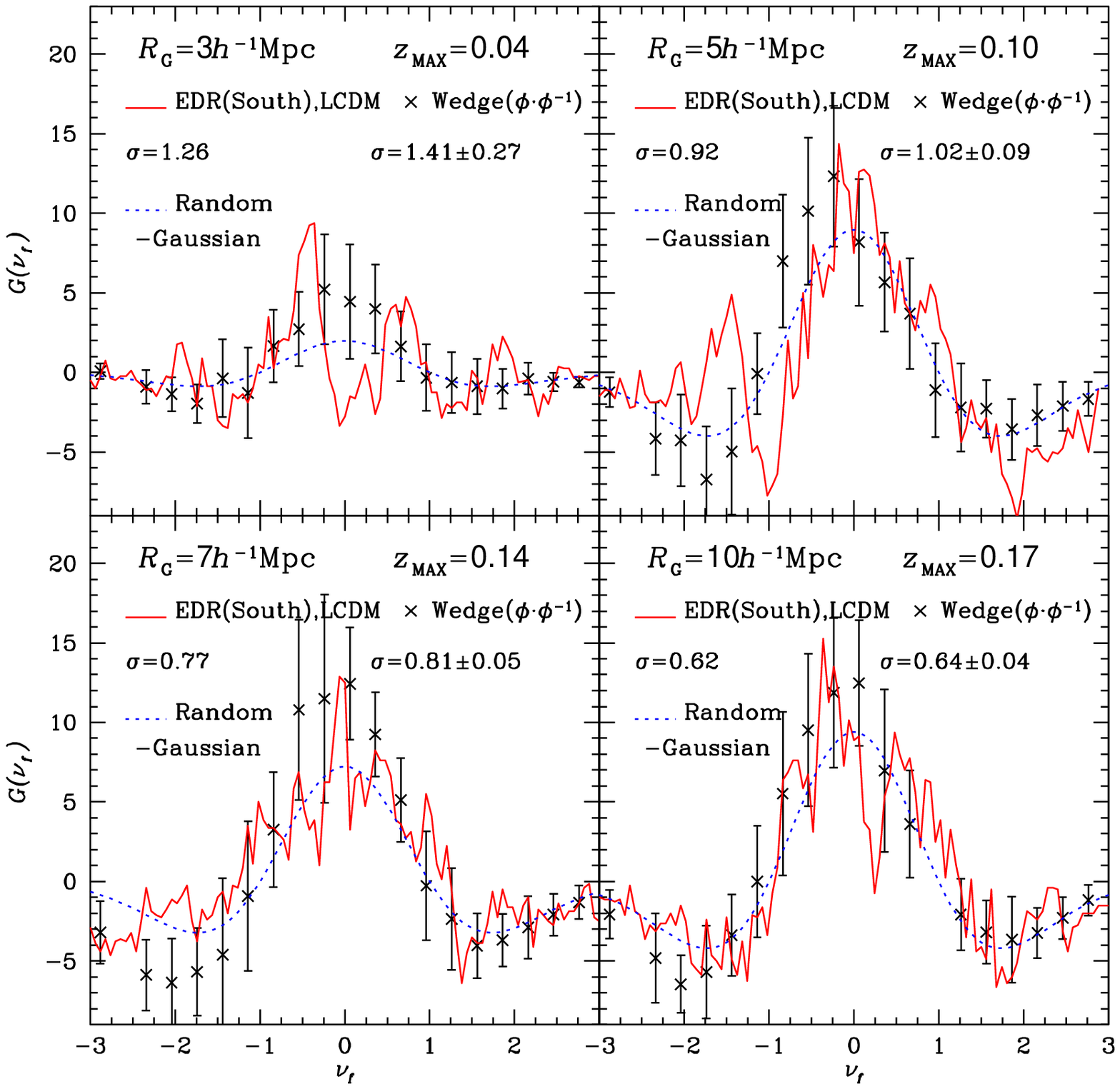}
\end{center}
\caption{Same as Figure \ref{fig:sdss_mock_LCDM} but plotted as a
 function of $\nuf$, the threshold defined through 
the volume fraction. The Gaussian prediction (eq.[\ref{eq:genus_rd}])
with its amplitude fitted to the SDSS genus are also plotted 
with dotted lines.}  
\label{fig:sdss_mock_LCDM_f}
\end{figure}

\begin{figure}[thp]
\begin{center}
\FigureFile(80mm,80mm){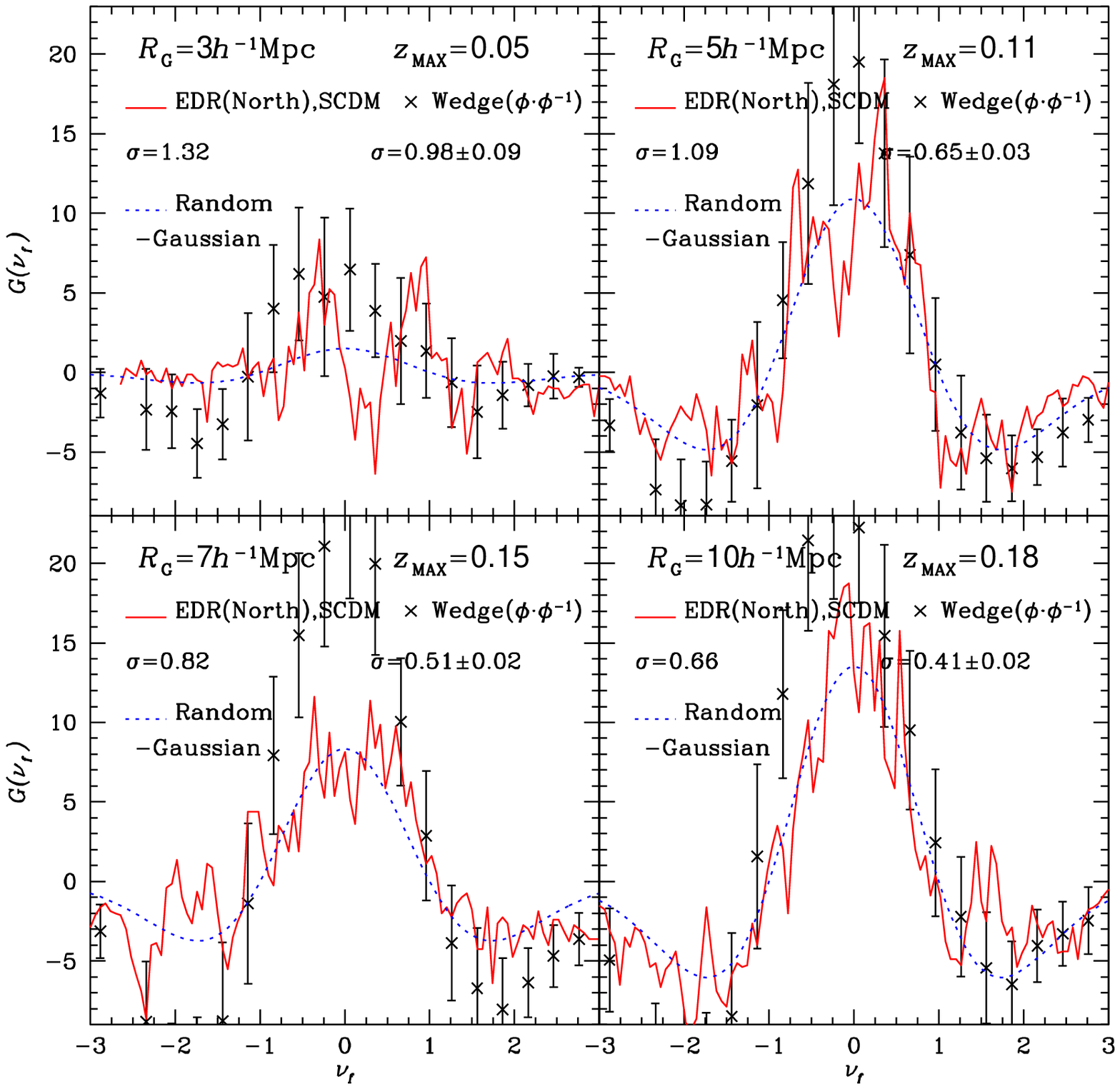}
\FigureFile(80mm,80mm){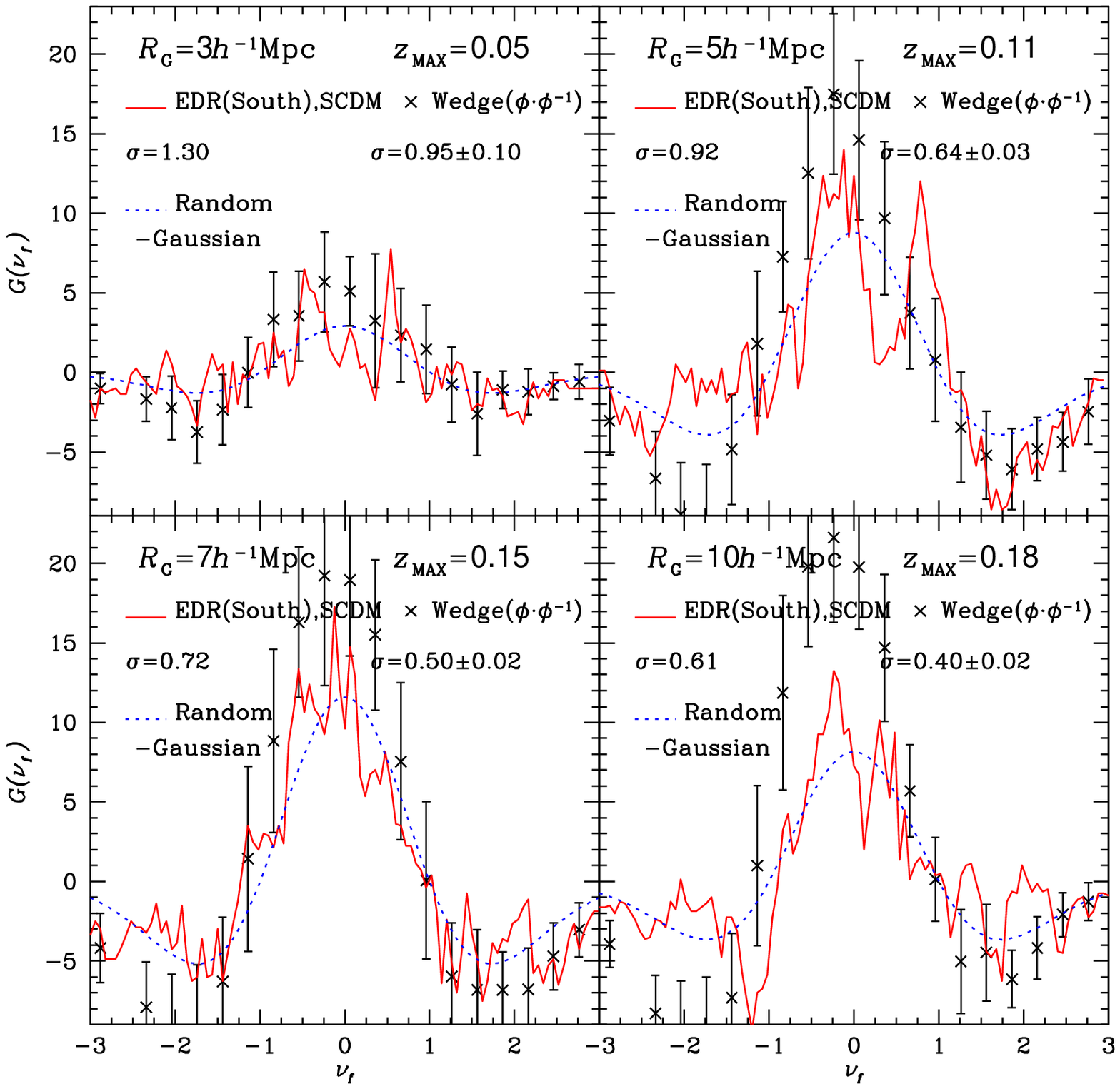}
\end{center}
\caption{Same as 
Figure \ref{fig:sdss_mock_LCDM_f} but in
SCDM model.}  
\label{fig:sdss_mock_SCDM_f}
\end{figure}

To allow clear comparison with other genus estimates in the
literature, in Figures \ref{fig:sdss_mock_LCDM_f} and
\ref{fig:sdss_mock_SCDM_f} we replot Figures \ref{fig:sdss_mock_LCDM}
and \ref{fig:sdss_mock_SCDM}, respectively, using $\nuf$. The solid
lines indicate the results from the SDSS EDR samples, and the dotted
curves show the Gaussian prediction (eq.[\ref{eq:genus_rd}]) with its
one free parameter, the amplitude, set by minimizing $\chi^2$ between
the SDSS genus and the theoretical curve.  The mock sample data are
shown as symbols with error bars to indicate the expected
sample-to-sample variation for the current data size.  Shown in this
way, the genus curves for both the data and mock samples appear to
agree with the random-Gaussian curve, which may be interpreted to
imply that the primordial Gaussianity is confirmed. 

A more correct interpretation is that, given the size of the estimated
uncertainties, these data do not provide evidence for initial
non-Gaussianity, i.e., the data are {\it consistent} with primordial
Gaussianity. In order to go further and place more quantitative
constraints on primordial Gaussianity with up-coming data, one needs a
more precise and reliable theoretical model for the genus which properly
describes the nonlinear gravitational effect possibly as well as galaxy
biasing beyond the simple mapping on the basis of the volume fraction
(see also Fig. \ref{fig:future} below).  A perturbative approach by
Matsubara (1994) combined with the extensive simulation mock sample
analysis may be a promising strategy for this purpose.


\section{Summary and Conclusions}

We present the first analysis of the three-dimensional genus statistics
for SDSS EDR galaxy data.  Due to the complicated survey volume and the
selection function of the current sample, analytic predictions of the
corresponding genus statistics are not feasible and we construct
extensive mock catalogs from N-body simulations in order to explore the
cosmological implications. We also use these mock catalogs to examine
the effects of several possible observational systematics on the genus
statistic.  We find that redshift-space distortions cause minimal
changes in the genus curve.  However, the slice-like geometry of the
SDSS EDR galaxy samples causes large distortions of the genus curve on
smoothing scales $R_{\rm G}> 7 h^{-1}$Mpc. We demonstrate that weighting
by the inverse of the selection function with distance in
apparent-magnitude limited samples allows recovery of the genus curve.
Because the mock catalogs include all of these effects, comparison with
theoretical models is possible.

We conclude that the observed shape and amplitude of the genus for the
SDSS EDR galaxy sample are consistent with the distribution of dark
matter particles in simulations of the LCDM model. In contrast, the
SCDM model predictions do not match the amplitude of the observed
genus.  Comparison of the observed genus curves with the theoretical
prediction for a Gaussian random field shows that, within the
uncertainties as estimated from the mock samples, the data are
consistent with Gaussianity of the primordial density field.

\begin{figure}[htp]
\begin{center}
\FigureFile(80mm,80mm){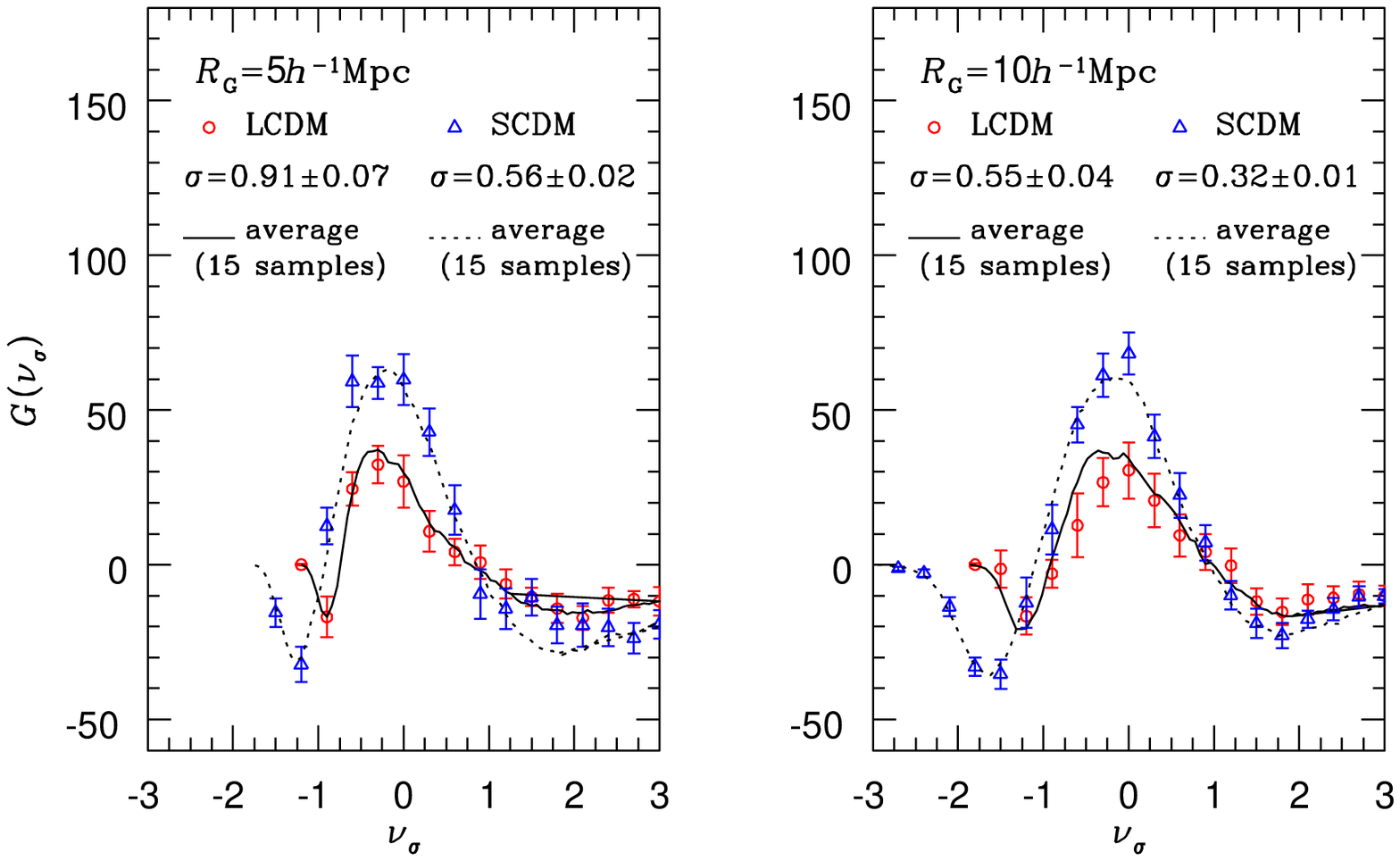}
\FigureFile(80mm,80mm){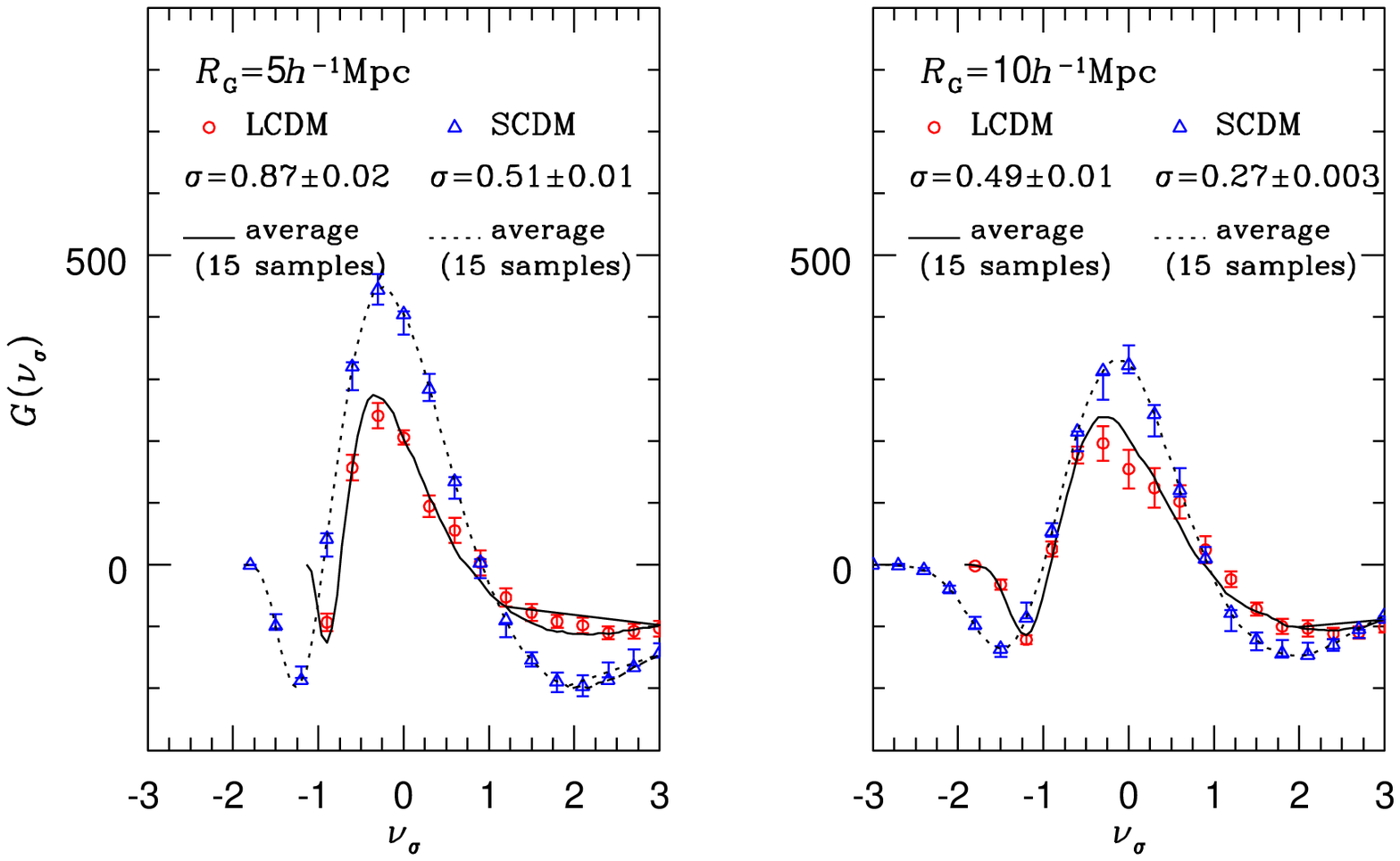}
\FigureFile(80mm,80mm){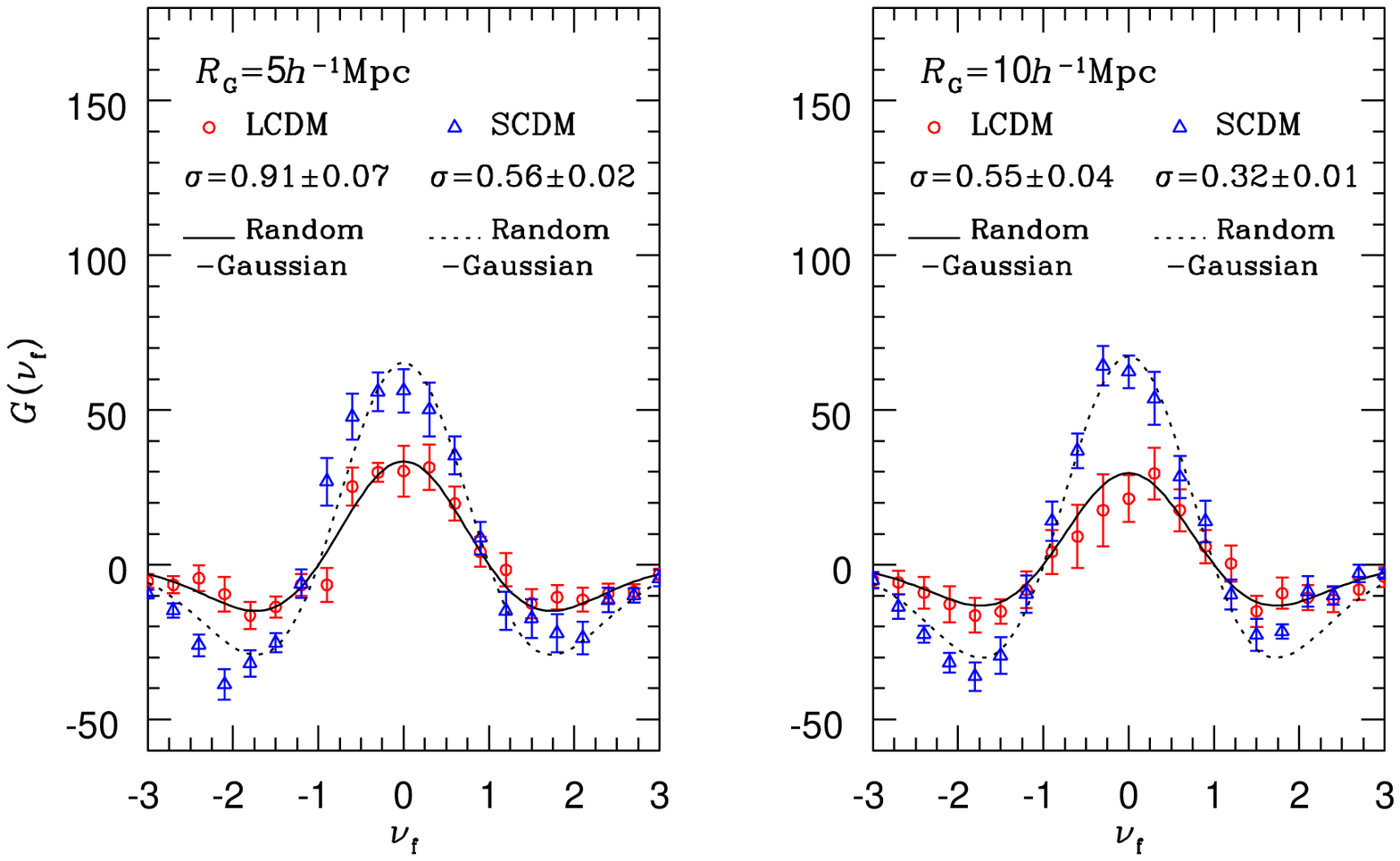}
\FigureFile(80mm,80mm){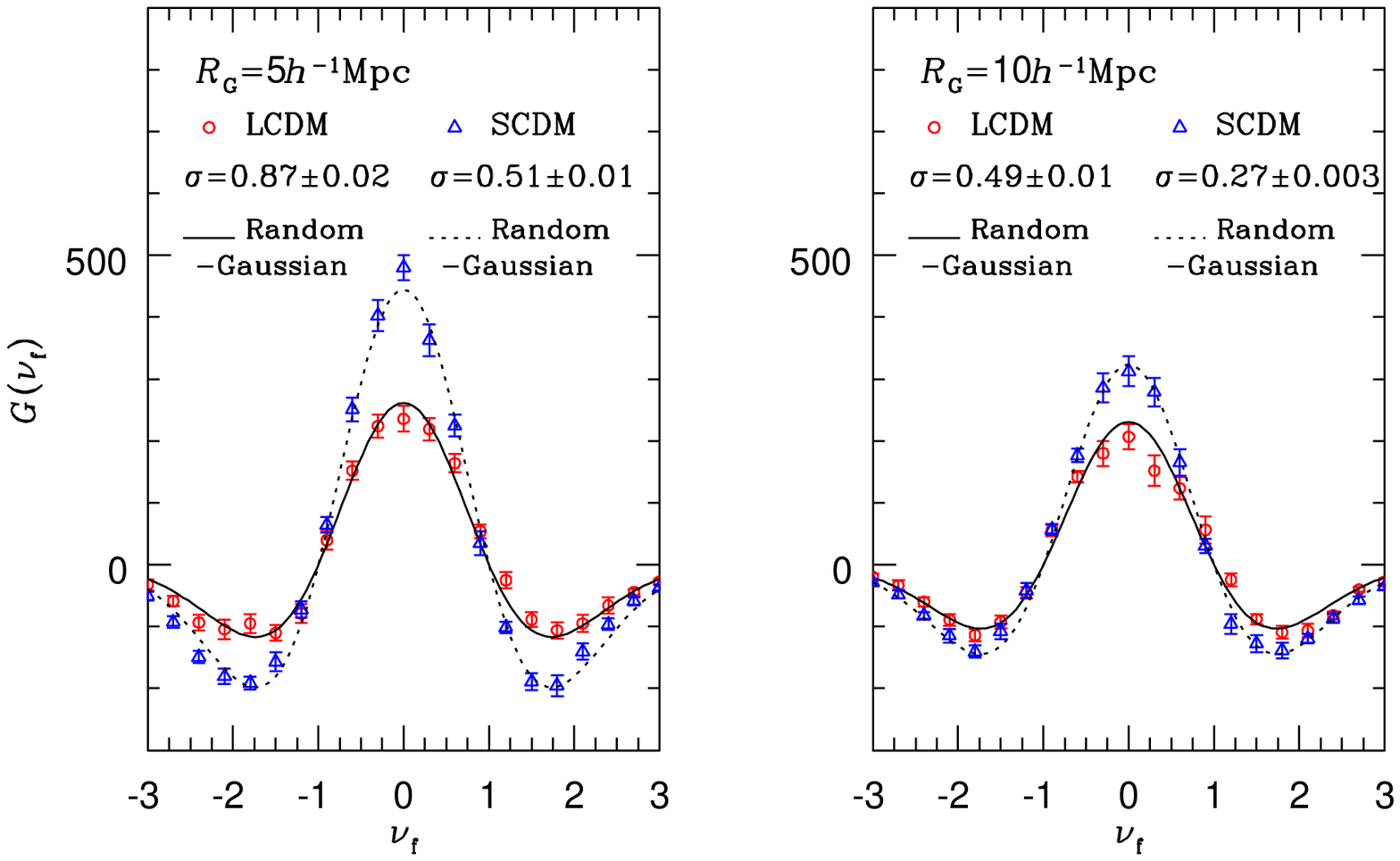}
\end{center}
\caption{Prediction of the genus for future samples of SDSS galaxies,
  using mock catalogs in LCDM ({\it Open circles} and {\it Solid lines})
  and SCDM ({\it Open triangles} and {\it Dotted lines}) From left to
  right, $R_{\rm G}=5h^{-1}$Mpc with $7 \times 10^4$ galaxies, $R_{\rm
  G}=10h^{-1}$Mpc with $7 \times 10^4$ galaxies, $R_{\rm G}=5h^{-1}$Mpc
  with $7 \times 10^5$ galaxies, and $R_{\rm G}=10h^{-1}$Mpc with $7
  \times 10^5$ galaxies.  Upper and lower panels plot the total number
  of genus in terms of $\nus$ and $\nuf$, respectively.}
  \label{fig:future}
\end{figure}

A question for the future is the degree of improvement that one can
expect for the larger samples of the SDSS galaxy redshift survey. To
predict this, we generate larger mock catalogs by increasing the sky
coverage by three times ($\sim 7\times 10^4$ galaxies) and thirty times
($\sim 7\times 10^5$ galaxies) relative to the Northern stripe in EDR.
Figure \ref{fig:future} shows the predicted genus in LCDM and SCDM
models (see also Colley et al.  2000 for predictions of genus results
for the SDSS). The positions of the symbols in those panels correspond
to the data of one specific mock sample, but with error-bars estimated
from the 15 independent mock samples. For reference, the lines in upper
panels show the mean genus curves averaged over the 15 independent mock
samples, while the lines in lower panels indicate the fit of the
random-Gaussian prediction to the plotted sample data.  Of interest is
that the smaller uncertainties in this future sample will put much
stronger constraints on cosmological models, the nonlinear nature of the
galaxy biasing, and primordial Gaussianity than those presented here.
It may also be worth while to apply 
theoretical estimate of genus statistics based on Lagrangian 
perturbation theory after appropriate smoothing (Seto et al. 1997).

\bigskip

We thank Y. P. Jing for kindly providing us his N-body simulation data
which were used in generating mock samples.  We also thank T. Buchert
for a careful reading of this manuscript.
I. K. gratefully
acknowledges support from the Takenaka-Ikueikai fellowship.  Numerical
computations were carried out at ADAC (the Astronomical Data Analysis
Center) of the National Astronomical Observatory, Japan.  This research
was supported in part by the Grant-in-Aid from Monbu-Kagakusho and Japan
Society of Promotion of Science (12640231, 13740150, 14102004, and
1470157). MSV and FH acknowledge support from NSF grant AST-0071201 and
a grant from the John Templeton Foundation. JRG acknowledges support
from NSF grant AST-9900772.

Funding for the creation and distribution of the SDSS Archive has been
provided by the Alfred P. Sloan Foundation, the Participating
Institutions, the National Aeronautics and Space Administration, the
National Science Foundation, the U.S. Department of Energy, the Japanese
Monbu-Kagakusho, and the Max Planck Society. The SDSS Web site is
http://www.sdss.org/.

The SDSS is managed by the Astrophysical Research Consortium (ARC) for
the Participating Institutions, which are the University of Chicago,
Fermilab, the Institute for Advanced Study, the Japan Participation
Group, the Johns Hopkins University, Los Alamos National Laboratory,
the Max-Planck-Institute for Astronomy (MPIA), the
Max-Planck-Institute for Astrophysics (MPA), New Mexico State
University, Princeton University, the United States Naval Observatory,
and the University of Washington.

\clearpage



\begin{thebibliography}{}
\bibitem[Adler(1981)]{A1981}
Adler,~R.~J. \ 1981, {\it The Geometry of Random Fields} 
(Chichester: Wiley)
\bibitem[Bardeen et al.(1986)]{BBKS1986}
Bardeen,~J.~M., Bond,~J.~R., Kaiser,~N., \& Szalay,~A.~S. \ 1986, 
\apj, 304, 15  
\bibitem[Barrow, Bhavsar \& Sonoda (1985)]{BBS1985}
Barrow,~J.~D., Bhavsar,~S.~P., \& Sonoda,~D.~H. \ 1985, \mnras, 216, 17
\bibitem[Blanton et al.(2001)]{B2001} 
Blanton,~M.~R., et al. \ 2001, \aj, 121, 2358 
\bibitem[Blanton et al.(2002)]{B2002}
Blanton,~M.~R., Lupton,~R.~H., Maley,~F.~M., Young,~N., Zehavi,~I., 
\& Loveday,~J. \ 2002, \aj, submitted
\bibitem[Canavezes et al.(1998)]{Canavezes98}
Canavezes,~A., et al. \ 1998, \mnras, 297, 777
\bibitem[Coles \& Jones(1991)]{CJ1991}
Coles,~P., \& Jones,~B. \ 1991, \mnras, 248, 1
\bibitem[Coles et al. 1991]{C1991}
Coles,~P., \& Plionis,~M. \ 1991, \mnras, 250, 75
\bibitem[Colley et al.(2000)]{CGWPB2000}
Colley,~W.~N., Gott,~J.~R.~III, Weinberg,~D.~H., Park,~C., 
\& Berlind,~A.~A. \ 2000, \apj, 529, 795
\bibitem[Connolly et al. (2002)]{C2002}
Connolly,~A.~J., et al. 2002, \apj, submitted, astro-ph/0107417
\bibitem[Dodelson et al. (2002)]{D2002}
Dodelson,~S., et al. 2002, \apj, submitted, astro-ph/0107421
\bibitem[Doroshkevich et al.(1970)]{D1970}
Doroshkevich,~A.~G. 1970, Astrofizika, 6, 581 (English transl. 
Astrophysics, 6, 320)
\bibitem[Eisenstein et al. (2001)]{E2001}
Eisenstein, D. J., et al. 2001, \aj, 122, 2267
\bibitem[Fukugita et al. (1995)]{FSI1995}
Fukugita,~M., Shimasaku,~K., \& Ichikawa,~T. 1995, \pasp, 107, 945
\bibitem[Fukugita et al. (1996)]{FIGDSS1996}
Fukugita,~M., Ichikawa,~T., Gunn,~J.~E., Doi,~M., Shimasaku,~K., \&
Schneider,~D.~P. \ 1996, \aj, 111, 1748
\bibitem[Gott, Mellot \& Dickinson(1986)]{GMD1986}
Gott,~J.~R.~III, Mellot,~A.~L., \& Dickinson,~M.\ 1986, \apj, 306, 341
\bibitem[Gott et al.(1989)]{G1989}
Gott,~J.~R.~III, et al.\ 1989, \apj, 340, 625 
\bibitem[Gunn et al.(1998)]{G1998}
Gunn,~J.~R.~III, et al. 1998, \apj, 116, 3040
\bibitem[Hamana, Colombi, \& Suto (2001)]{HCS2001}
Hamana,~T., Colombi,~S., \& Suto,~Y. 2001, A\&A, 367, 18
\bibitem[Hamilton et al.(1986)]{HGW1986}
Hamilton,~A.~J.~S., Gott,~J.~R.~III, \& Weinberg,~D.~H.\ 1986, 
\apj, 309, 1
\bibitem[Hamilton et al.(1991)]{HMKL1991} 
Hamilton,~A.~J.~S., Matthews,~A., Kumar,~P., \& Lu, E.\ 1991, 
\apjl, 374, L1 
\bibitem[Hikage, Taruya \& Suto(2001)]{HTS2001}
Hikage,~C., Taruya,~A., \& Suto,~Y.\ 2001, \apj, 556, 641
\bibitem[Hikage, Taruya \& Suto(2002)]{HTS2002}
Hikage,~C., Taruya,~A., \& Suto,~Y.\ 2002, \pasj, submitted
\bibitem[Hogg et al.(2001)]{Hogg2001}
Hogg,~D.~W., Finkbeiner,~D.~P., Schlegel,~D.~J., \& Gunn,~J.~E. 2001, 
\aj, 122, 2129
\bibitem[Hoyle et al. (2002a)]{H2002a}
Hoyle,~F., Vogeley,~M.~S., \& Gott,~J.~R.~III, 2002a, \apj, 570, 44
\bibitem[Hoyle et al.(2002b)]{H2002b}
Hoyle,~F., Vogeley,~M.~S., Gott,~J.~R.~III, Blanton,~M., Tegmark,~M., 
Weinberg,~D.~H, Bahcall,~N., Brinkmann,~J., \& York,~D.~G 2002b, \apj,
submitted, astro-ph/0206146
\bibitem[Infante et al. (2002)]{I2002}
Infante,~L., et al. 2002, \apj, 567, 1551
\bibitem[Jing \& Suto (1998)]{JS1998}
Jing,~Y.~P., \& Suto,~Y. 1998, \apj, 494, L5
\bibitem[Kayo et al.(2001)]{KTS2001}
Kayo,~I., Taruya,~A., \& Suto,~Y.\ 2001, \apj, 561, 22
\bibitem[Kayo et al.(2002)]{KTS2002}
Kayo,~I., et al. 2002, \apj, submitted
\bibitem[Lupton (2002)]{L2002}
Lupton,~R.~H. 2002, in preparation
\bibitem[Matsubara(1994)]{M1994}
Matsubara,~T.\ 1994, \apjl, 434, 43
\bibitem[Matsubara(1996)]{M1996} 
Matsubara,~T.\ 1996, \apj, 457, 13. 
\bibitem[Matsubara \& Suto(1996)]{MS1996}
Matsubara,~T., \& Suto,~Y.\ 1996, \apj, 460, 51
\bibitem[Matsubara \& Yokoyama(1996)]{MY1996}
Matsubara,~T., \& Yokoyama,~J.\ 1996, \apj, 463, 409
\bibitem[Mecke, Buchert \& Wagoner(1994)]{MBW1994}
Mecke,~K.~R., Buchert,~T. \& Wagner,~H. \ 1994, A\&A, 288, 697
\bibitem[Melott (1987)]{M1987} 
Melott,~A.~L. 1987, Topology of the Universe: Motivation
 for the Study of Large -Scale Structure, in Proceedings of the XIIIth
 Texas Symposium on Relativistic Astrophysics, M. Elmer ed, 1987, World
 Scientific Publishing
\bibitem[Melott \& Dominik (1993)]{MD1993}
Melott,~A.~L., \& Dominik,~K.\ 1993, \apjs, 86, 1
\bibitem[Moore et al. (1992)]{MFWSLEKER1992}
Moore,~B., Frenk,~C.~S., Weinberg,~D.~H., Saunder,~W., Lawrence,~A.,
Ellis,~R.~S., Kaiser,~N., Efstathiou,~G., \& Rowan-Robinson,~M.
\ 1992, \mnras, 256, 477
\bibitem[Park, Gott, \& Choi (2001)]{PGC2001}
Park,~C., Gott,~J.~R.~III, \& Choi,~Y.~J. 2001, \apj, 553, 33
\bibitem[Park, Gott, \& da Costa (1992)]{Park1992a}
Park,~C., Gott,~J.~R.~III, \& da Costa,~L.~N. 1992, \apj, 392, L51
\bibitem[Park et al. (1992)]{Park1992b}
Park,~C., Gott,~J.~R.~III, Melott,~A.~L., \& Karachentsev,~I.~D. 1992, 
\apj, 387, 1
\bibitem[Peacock \& Dodds(1996)]{PD1996}
Peacock,~J.~A., \& Dodds,~S.~J. 1996, \mnras, 280, L19
\bibitem[Peebles (1980)]{Peebles1980}
Peebles,~P.~J.~E. 1980, The Large-Scale Structure of the Universe
 (Princeton: Princeton University Press)
\bibitem[Pier et al. (2002)]{P2002}
Pier,~J., et al. 2002, \aj, submitted
\bibitem[Plionis, Valdarnini, \& Coles (1992)]{PVC1992}
Plionis,~M., Valdarnini,~R., \& Coles,~P. 1992, \mnras, 258, 114
\bibitem[Rhoads et al.(1994)]{RGP1994}
Rhoads,~J.~E.~R., Gott,~J.~R.~III, \& Postman,~M. 1994, \apj, 421, 
\bibitem[Schmalzing \& Buchert(1997)]{SB1997}
Schmalzing,~J., \& Buchert,~T.\ 1997, \apj, 482, L1
\bibitem[Shandarin(1983)]{Shandarin}
Shandarin,~S.~F. 1983, Sov.Astron.Lett. 9, 104
\bibitem[Schlegel, Finkbbeiner, \& Davis (1998)]{SFD98}
Schlegel,~D.~J., Finkbeiner,~D.~P., \& Davis,~M. 1998, \apj, 500, 525
\bibitem[Seto et al. (1997)]{Seto1997}
Seto,~N., Yokoyama,~J., Matsubara,~T., \& Siino,~M. 1997, \apjs, 110, 177
\bibitem[Smith et al.(2002)]{Smith2002}
Smith,~J.~A., et al. 2002, \aj, 123, 2121.
\bibitem[Stoughton et al. (2002)]{Stoughton2002}
Stoughton,~C., et al. 2002, \aj, 123, 485
\bibitem[Strauss et al. (2002)]{Strauss2002} 
Strauss,~M.~A., et al. 2002, \apj, in press
\bibitem[Szalay et al. (2002)]{Szalay2002}
Szalay,~A.~S., et al. 2002, \apj, submitted, astro-ph/0107419
\bibitem[Szapudi et al. (2002)]{Szapudi2002}
Szapudi,~I., et al. 2002, \apj, submitted, astro-ph/0111058
\bibitem[Tegmark et al. (2002)]{Tegmark2002}
Tegmark,~M., et al. 2002, \apj, 571, 191
\bibitem[Totsuji \& Kihara(1969)]{TK69}
Totsuji,~H., \& Kihara,~T.\ 1969, \pasj, 21, 221
\bibitem[Vogeley et al.(1994)]{VPGHG1994}
Vogeley,~M.~S., Park,~C., Geller,~M.~J., Huchra,~J.~P., 
\& Gott,~J.~R.~III  1994, \apj, 420, 525
\bibitem[Weinberg (1988)]{W1988} 
Weinberg,~D.~H. 1988, \pasp, 100, 1373
\bibitem[White (1979)]{White1979}
White,~S.~D.~M. 1979, \mnras, 186, 145
\bibitem[York et al.(2000)]{Y2000} 
York,~D.~G., et al.\ 2000, \aj, 120, 1579. 
\bibitem[Zehave et al. (2002)]{Zehavi2002}
Zehavi,~I., et al. 2002, \apj, 571, 172
\end{thebibliography}
\end{document}